\documentclass[9pt,twocolumn,twoside]{osajnl}
\journal{jocn} 

\usepackage{epsfig,amsmath,amssymb,epsf,amsthm,scalefnt,multirow,multicol}
\usepackage{amsmath}
\usepackage{xcolor}
\usepackage{float}
\usepackage{psfrag}
\usepackage{graphicx}
\usepackage{subcaption}
\newsavebox{\imagebox}

\newtheorem*{lemma*}{Lemma}

\def\b0{{\pmb{0}}} 

\def\ba{{\mathbf{a}}}   
  \def\bg{{\mathbf{g}}} \def\bh{{\mathbf{h}}}

 \def\br{{\mathbf{r}}} \def\bs{{\mathbf{s}}} 
  \def\bw{{\mathbf{w}}} \def\bx{{\mathbf{x}}}

\def\bA{{\mathbf{A}}}   
  \def\bG{{\mathbf{G}}} \def\bH{{\mathbf{H}}}
\def\bI{{\mathbf{I}}} \def\bJ{{\mathbf{J}}}  \def\bL{{\mathbf{L}}}
\def\bM{{\mathbf{M}}} \def\bN{{\mathbf{N}}}  
   \def\bT{{\mathbf{T}}}
  \def\bW{{\mathbf{W}}}

\setboolean{shortarticle}{false}

\title{On the Physical Layer Security of Visible Light Communications Empowered by Gold Nanoparticles}

\author[1]{Geonho Han}
\author[1]{Hyuckjin Choi}
\author[2]{Ryeong Myeong Kim}
\author[2]{Ki Tae Nam}
\author[1*]{Junil Choi}
\author[3]{Theodoros A. Tsiftsis}

\affil[1]{School of Electrical Engineering, Korea Advanced Institute of Science and Technology, Daejeon 34141, South Korea}
\affil[2]{Department of Materials Science and Engineering, Seoul National University, Seoul 08826, South Korea}
\affil[3]{Department of Informatics and Telecommunications, University of Thessaly, 35100 Lamia, Greece}

\affil[*]{Corresponding author: junil@kaist.ac.kr}

\begin{abstract}
Visible light is a proper spectrum for secure wireless communications because of its high directivity and impermeability in indoor scenarios. However, if an eavesdropper is located very close to a legitimate receiver, secure communications become highly risky. In this paper, to further increase the level of security of visible light communication (VLC) and increase its resilience against to malicious attacks, we propose to capitalize on the recently synthesized gold nanoparticles (GNPs) with chiroptical properties for circularly polarized light resulting the phase retardation that interacts with the linear polarizer angle. GNP plates made by judiciously stacking many GNPs perform as physical secret keys. Transmitters send both the intended symbol and artificial noise to exploit the channel variation effect by the GNP plates, which is highly effective when an eavesdropper is closely located to the legitimate receiver. A new VLC channel model is first developed by representing the effect of GNP plates and linear polarizers in the circular polarization domain. Based on the new channel model, the angles of linear polarizers at the transmitters and legitimate receiver are optimized considering the effect of GNP plates to increase the secrecy rate in wiretapping scenarios. Simulations verify that when the transmitters are equipped with GNP plates, even if the eavesdropper is located right next to the legitimate receiver, insightful results on the physical layer security metrics are gained as follows: 1) the secrecy rate is significantly improved and 2) the symbol error rate gap between the legitimate receiver and eavesdropper becomes much larger due to the chiroptical properties of GNP plates.
\end{abstract}

\setboolean{displaycopyright}{false} 

\begin{document}

\maketitle

\section{Introduction}\label{sec_intro}
Future wireless communication systems require high data rate, ultra-reliability, and low latency to satisfy the needs of advanced applications while available spectrum is limited. For short range communications, one of the candidates to fulfill these requirements is a visible light communication (VLC) system that usually exploits light emitting diodes (LEDs) as transmitters, and photodiodes (PDs) as receivers \cite{pathak:2015,jovicic:2013}. In VLC systems, the shortage of spectrum is not an issue due to the extremely high frequency, i.e., about 400 to 790 THz, of visible light. Moreover, the inherent directivity and impermeability of VLC signals prevent information leakage, which enhance the secrecy rate compared to the wireless communication systems using radio frequency (RF) bands \cite{Zvanovec:2015,Sadat:2022}. Due to these advantages, beyond fifth generation (5G) or sixth generation (6G) networks aim to achieve high physical layer security by adopting the visible spectrum \cite{Osorio:2022,Solaija:2022,Irram:2022}. There are additional benefits for the VLC systems such as low implementation cost by using existing infrastructure, and dual function of illumination/communications with low power consumption using LEDs \cite{Karunatilaka:2015}.

\subsection{Technical Literature Review}
Various works have been studied for precoding techniques in VLC systems to enhance the physical layer security. In \cite{Mostafa:2014}, the artificial noise was added to zero-forcing (ZF) precoder to improve the secrecy rate. A signal-to-interference-plus-noise ratio (SINR) maximization problem was performed by optimizing simultaneous beamforming and jamming to avoid wiretapping with a randomly located eavesdropper (Eve) in \cite{Cho:2019}. In \cite{Cho:2020}, a ZF beamformer was designed to minimize secrecy outage probability with the signal-to-noise ratio (SNR) constraint for a legitimate receiver (Bob) by sampling the SNR space. A robust precoding for VLC systems using imperfect location information of Eve was proposed to guarantee the secrecy rate in \cite{Mostafa:2015}. In \cite{Mostafa:2016}, the transmit beamformers were designed to maximize the achievable secrecy rate considering the imperfect channel information, and channel mismatch. Robust secure beamforming was developed in simultaneous lightwave information and power transfer (SLIPT) VLC networks to support energy-limited users with a nonlinear energy harvesting model in \cite{Liu:2020SLIPT}. A hybrid system combining RF and VLC was proposed to maximize the sum secrecy rate of multi-users using joint precoding and user association in \cite{Da:2023}. In \cite{Albayrak:2023}, the inter-symbol interference generated by reflections was compensated by the ZF and artificial noise beamformers with the knowledge of Eve's channel state information (CSI). In \cite{Pan:2020,Pan:2022,Pan:2021}, quantum key distribution (QKD) protocols led to high secret key rates by considering limitations on Eve's power collection ability and implementing exclusion zones around Bob.

Recently, intelligent reflecting surface (IRS) has been used in VLC systems to improve security by exploiting the advantages of IRS, e.g., the power of creating favorable multi-paths, low power consumption, and coverage extension. An IRS-aided VLC system was designed to improve the physical layer security by making the signals toward Eve weak using the IRS reflections in \cite{Sun:2021}. In \cite{Sun:2022}, an IRS was adopted to increase the secrecy rate with an iterative Kuhn-Munkres algorithm in a multi-user situation. In \cite{Saifaldeen:2022}, the beamforming weights at LEDs and IRSs were optimized to achieve high secrecy capacity by using deep reinforcement learning.

There also have been many works to improve the secrecy rate of VLC systems by developing modulation schemes. In \cite{Su:2021}, a constellation design based on generalized space shift keying (GSSK) was proposed to minimize the bit error rates (BERs) of legitimate receivers simultaneously degrading Eve's BER performance. Secrecy rate bounds were derived in \cite{Wang:2019} to better exploit spatial modulation that activates one transmitter at each time. A novel modulation technique modified from multiple pulse position modulation (MPPM) was proposed in \cite{Sejan:2022} to provide secure Internet of Things (IoT) connectivity. In \cite{Nie:2022}, pairwise coding (PC)-based multiband carrierless amplitude and phase (mCAP) modulation with chaotic dual-mode index modulation (DM) was developed for secure VLC systems.

\subsection{Motivation}
Most of previous works did not put much attention to the polarization of visible light though. In \cite{Yeh:2015,Yang:2021,Wang:2014,De:2020}, linear polarizers were used to generate the orthogonally polarized light with either $0^{\circ}$ or $90^{\circ}$ degrees to mitigate mutual interference. In \cite{Ivanovich:2018}, polarization division multiplexing (PDM) was developed by representing the polarization states as the Stokes vectors and detecting the polarization using polarimeters. In \cite{Turke:2023}, a hybrid wavelength division multiplexing (WDM) and PDM system was suggested to achieve high capacity by using compressed return to zero advanced modulation.

Motivated by the advancements for PDM discussed in \cite{Ivanovich:2018,Turke:2023}, which increased the achievable rates, we propose judiciously exploiting the polarization of visible light to enhance the physical layer security. The detection of polarization using polarimeters in \cite{Ivanovich:2018}, however, is not appropriate for wireless communications since the rotation rate of waveplate in the polarimeter is too slow compared to a symbol period \cite{Schaefer:2007}. For practicality, we need to design a secure VLC system incorporating the polarization by only adopting PDs as the receivers.

\subsection{Our Contributions}
In this paper, we propose a gold nanoparticle (GNP)-empowered secure VLC system with linear polarizers for indoor environments. The GNPs have chiroptical properties that can yield different circular dichroism (CD) and optical rotatory dispersion (ORD) for the incident light, which denote the effects of differential absorption and refraction for left circularly polarized (LCP) and right circularly polarized (RCP) light, by controlling the type, pattern, and size of GNPs \cite{Kim:2021}.

The main contributions of our paper are summarized as follows.
\begin{itemize}
	\item We first derive the effects of GNP plate, which is a panel composed of GNPs, and linear polarizer in the circular polarization (CP) domain.	
	\item The overall VLC channel model including the effects of GNP plates and linear polarizers is derived when each transmitter is equipped with a GNP plate, a linear polarizer, and an LED, and a receiver is equipped with a linear polarizer and a PD.
	\item The angles of linear polarizers at the transmitters and Bob are optimized to improve the secrecy rates without the knowledge of Eve in indoor eavesdropping scenarios.
	\item It is numerically verified that the proposed GNP-empowered VLC system can significantly improve the secrecy rate, and symbol error rate (SER) metrics in particular for the locations near Bob.
\end{itemize}

Since this is the very first work proposing to exploit the GNPs for VLC systems, our work is corroborated by simulation results following the same steps as the development of other wireless communication technologies, e.g., multiple-input multiple-output (MIMO) systems \cite{Love:2003,Love:2003equal}, IRSs \cite{Wu:2019,Cui:2019,Zheng:2019}, or satellite communications \cite{You:2020,Talgat:2020}. We plan to perform real experiments to physically verify the effectiveness of proposed GNP-empowered VLC system in the near future.

Note that, although the GNPs require gold to fabricate, the cost can be quite affordable. For a GNP plate with $1$ cm $\times$ $1$ cm size and hexagonal pattern, the total volume of GNPs in the plate can be obtained as $V_{\mathrm{tot}}=\frac{A_{\mathrm{tot}}}{A_{\mathrm{hex}}}N_{\mathrm{hex}}V_{\mathrm{GNP}}$ where $A_{\mathrm{tot}}=1\,\mathrm{cm}^2$ is the total area of the plate, $A_{\mathrm{hex}}=12\sqrt{3}\times 10^{-14}\,\mathrm{m}^2$ is the area of a hexagon, $N_{\mathrm{hex}}=3$ is the number of GNPs in a hexagon, and $V_{\mathrm{GNP}}=(200\,\mathrm{nm})^3$ is the volume of a GNP. Using $V_{\mathrm{tot}}$, the density of gold, and the price of gold per one gram, this GNP plate production cost becomes just 1.17 cents. Therefore, the mass production of GNPs would be possible, making the GNP-empowered VLC system feasible. For more details of the fabrication of GNP plate, we refer to the section ``Methods" in \cite{Kim:2022}, which specifically elaborates the synthesis of GNPs and fabrication process of GNP plate.

\begin{figure*}
    \centering
    \includegraphics[width=1.4\columnwidth]{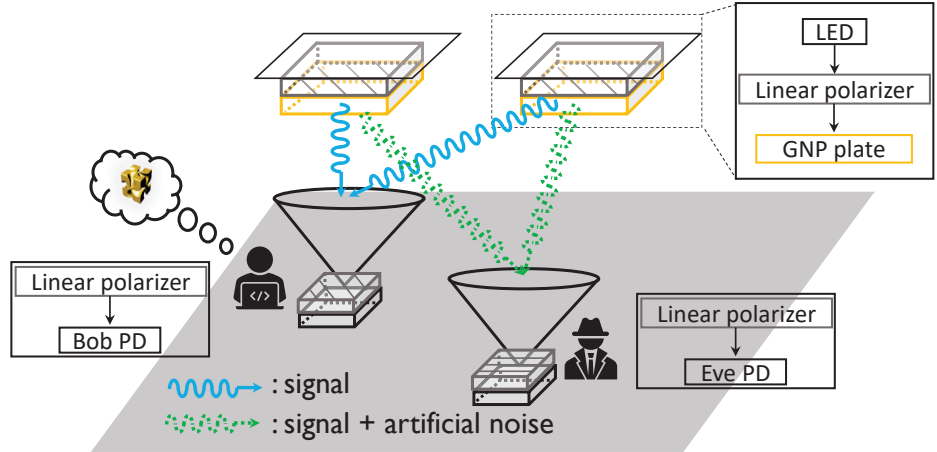}
    \caption{The conceptual figure of proposed GNP-empowered VLC system.}\label{sec_model}
\end{figure*}

This paper is organized as follows. In Section \ref{sec_II}, we first show the overall structure of GNP-empowered VLC system and present the chiroptical properties of GNPs and then derive the effective VLC channel model of GNP-empowered system. Taking this model into account, we optimize the linear polarizer angles at the transmitters and Bob to improve the performance and design the precoders for intended symbol and artificial noise in Section \ref{prec_angle}. In Section \ref{simul}, we show the performance improvement with respect to the secrecy rate and SER when using GNP-empowered VLC system in indoor scenarios via simulation results. Finally, we conclude the paper in Section \ref{conc}.

\textit{Notation:} $a$, $\ba$, and $\bA$ denote a scalar, vector, and matrix. $|a|$ and $\angle a$ denote the magnitude and angle of $a$. $\|\ba\|$ represents the Euclidean norm of $\ba$. $(\cdot)^{\mathrm{T}}$, $(\cdot)^*$, $(\cdot)^{-1}$, and $(\cdot)^{\dagger}$ are the transpose, complex conjugate, inverse, and pseudo inverse. A normal distribution with mean $\mu$ and variance $\sigma^2$ is denoted as $\mathcal{N}(\mu,\sigma^2)$. $\mathbb{R}_+$ represents the set of positive real numbers. $\mathcal{\mathbf{1}}_N$, $\mathbf{0}_{M\times N}$, and $\bI_N$ denote the $N\times 1$ all-ones vector, all-zeros matrix with the size $M\times N$, and identity matrix with the size $N\times N$. $\mathrm{Re}(\cdot)$ and $\mathrm{Im}(\cdot)$ are the operations extracting the real and imaginary parts of a given complex number. The Kronecker and Hadamard products of $\ba$ and $\mathbf{b}$ are represented as $\ba\otimes\mathbf{b}$ and $\ba\odot\mathbf{b}$. $\mathrm{max}(a,b)$ is the max function that returns the largest value between $a$ and $b$. $\mathrm{diag}(\cdot)$ denotes the diagonal matrix. $\mathbb{E}\left\{\cdot\right\}$ and $\mathcal{N}(\cdot)$ are the expectation and null space operators. $f'(\cdot)$ and $f''(\cdot)$ denote the first and second derivatives of a function $f(\cdot)$.

\begin{figure}
	\centering
	\includegraphics[width=1.0\columnwidth]{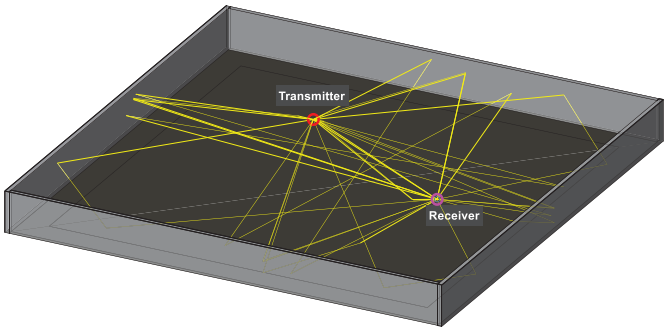}
	\caption{The VLC channel paths between a transmitter and a receiver generated by WiThRay \cite{Choi:2023}.}\label{WiThRay_ch}
\end{figure}

\section{System Model}\label{sec_II}
We focus on an indoor GNP-empowered VLC system with $N_t$ transmitters located separately, the legitimate receiver Bob, and a randomly located eavesdropper Eve.\footnote{While we focus only on the single Eve case in this section for simplicity, we consider multiple Eves for numerical studies in Section \ref{simul}.} The transmitters are uniformly located on the ceiling, and each transmitter is equipped with a GNP plate, a linear polarizer, and an LED, while Bob and Eve are both equipped with only a linear polarizer and a PD. Fig. \ref{sec_model} shows the concept of proposed GNP-empowered VLC system. The GNP plates work as physical secret keys; therefore, the proposed system is a hybrid scheme of using both artificial noise and physical secret key to enhance secrecy performance.

Due to the strong directivity of signals in the visible light spectrum, the line-of-sight (LOS) path has larger gain compared to the strongest non-LOS (NLOS) path, i.e., the first reflected path \cite{Komine:2004,Lee:2011,Mapunda:2020}. In \cite{Aboagye:2022}, the average sum rate of VLC system with the LOS blockage is sharply declined. We, however, take into account the NLOS paths as well to consider a realistic VLC channel, which is generated using a versatile ray-tracing simulator (WiThRay) \cite{Choi:2023}. The VLC channel paths between a transmitter and a receiver are depicted in Fig. \ref{WiThRay_ch}. The parameters used throughout this paper are listed in TABLE~I.

\begin{table}
	\centering
	\caption{List of important parameters.}
	\label{t1}
	\begin{tabular}{|c|l|}
		\noalign{\smallskip}\noalign{\smallskip}\hline
		Parameter & Description \\
		\hline
		$g_{mn}$ & the $n$-th path-loss of VLC channel \\
		& between the $m$-th LED and a PD \\
		$I_{\mathrm{DC}}$ & DC bias\\
		$a_{\{L,R\}}$ & absorption factors for LCP and RCP light  \\
		$\bar{a}_{\{L,R\}}$ & transmittance for LCP and RCP light\\
		$\Delta\varphi$  & difference of phase retardation \\
		& between LCP and RCP light\\
		$\theta_m$ & linear polarizer angle at the $m$-th transmitter \\
		$\theta_{\{B,E\}}$ & linear polarizer angles at Bob/Eve \\
		$\boldsymbol{\vartheta}_{\{B,E\},n}$ & phases of the $n$-th path\\
		& due to the propagation delay for Bob/Eve \\
		$\zeta$ & current-to-light conversion efficiency \\
		$\eta$ & PD's responsivity \\
		$(s_{\mathrm{I}},\bs_{\mathrm{a}})$ & intended symbol and artificial noise \\
		$(\bw,\bW_{\mathrm{a}})$ & precoders for $s$ and $\bs_{\mathrm{a}}$ \\
		\hline
	\end{tabular}
\end{table}

\subsection{GNP Properties}\label{sec_III}
\begin{figure}
	\centering
	\begin{subfigure}{0.5\linewidth}
		\centering
		\includegraphics[width=\linewidth]{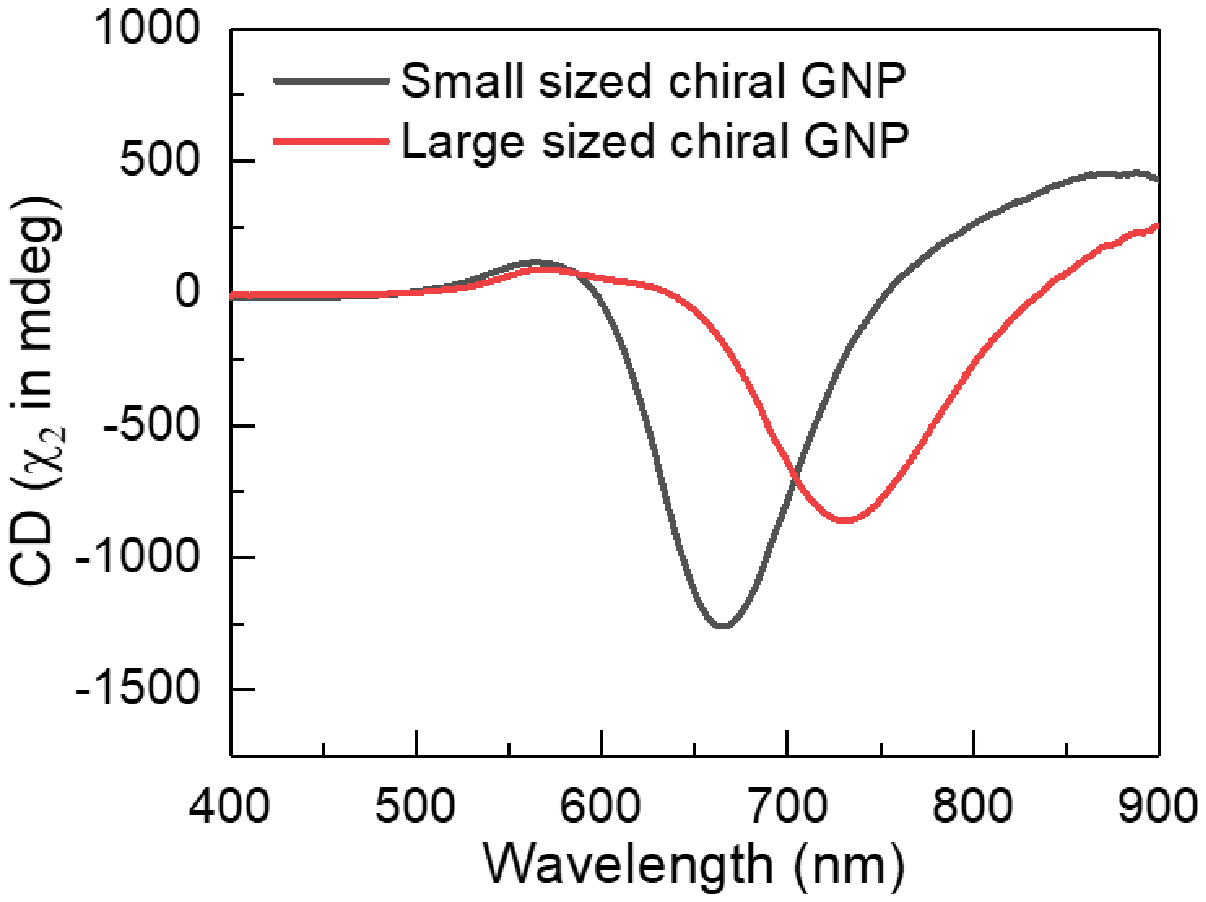}
		\label{CD_s}
	\end{subfigure}\hfill
	\begin{subfigure}{0.5\linewidth}
		\centering
		\includegraphics[width=\linewidth]{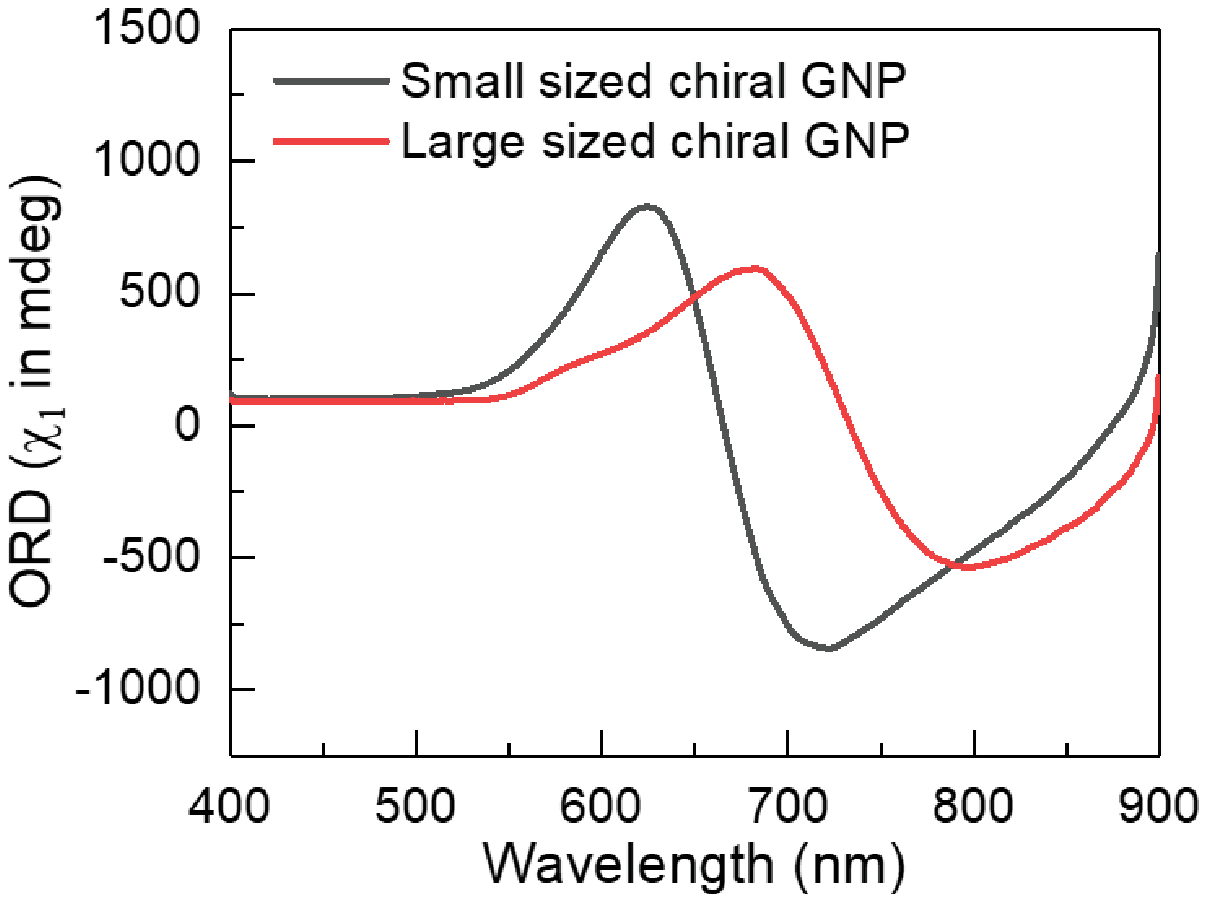}
		\label{ORD_s}
	\end{subfigure}
	
	\caption{The chiroptical properties of GNP plates depending on the different sizes with the same pattern.}\label{size}
\end{figure}

\begin{figure}
	\centering
	\begin{subfigure}{0.5\linewidth}
		\centering
		\includegraphics[width=\linewidth]{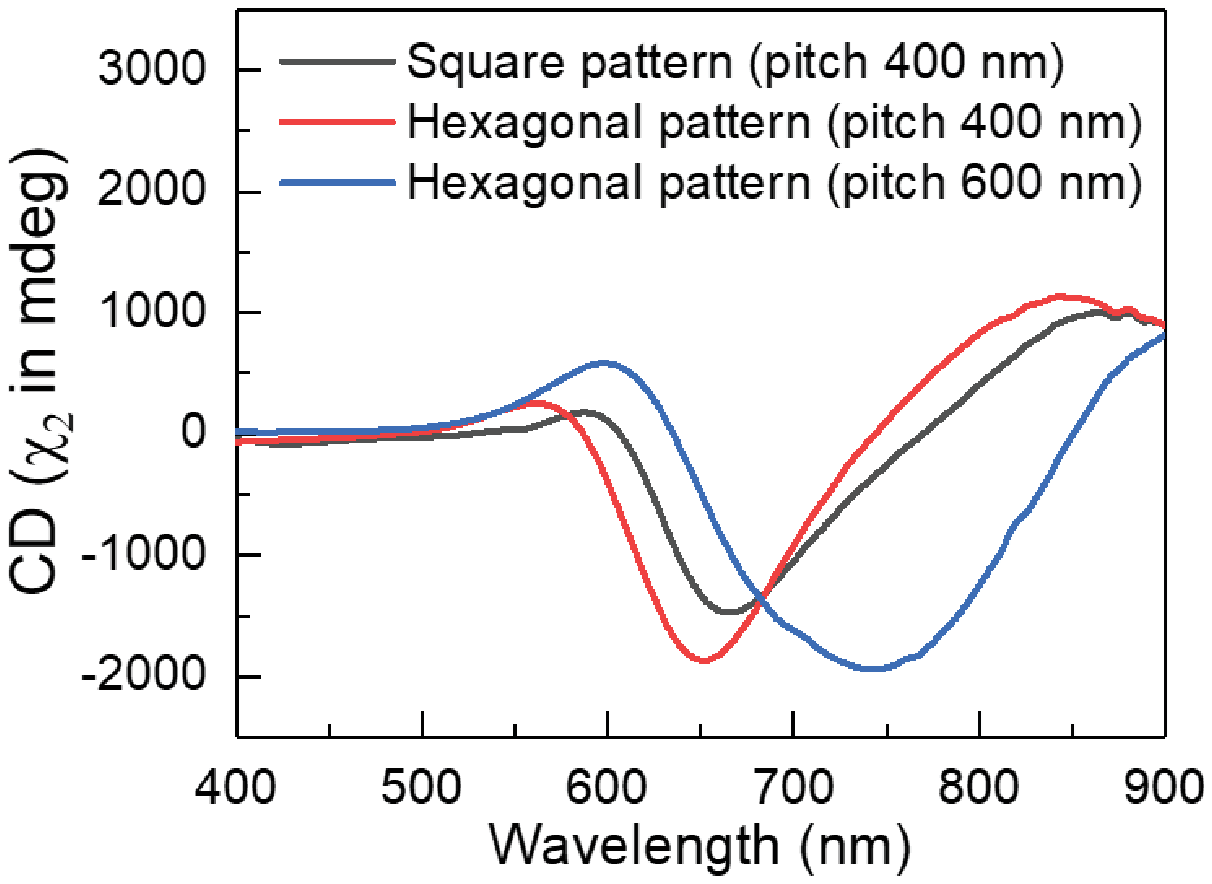}
		\label{CD_p}
	\end{subfigure}\hfill
	\begin{subfigure}{0.5\linewidth}
		\centering
		\includegraphics[width=\linewidth]{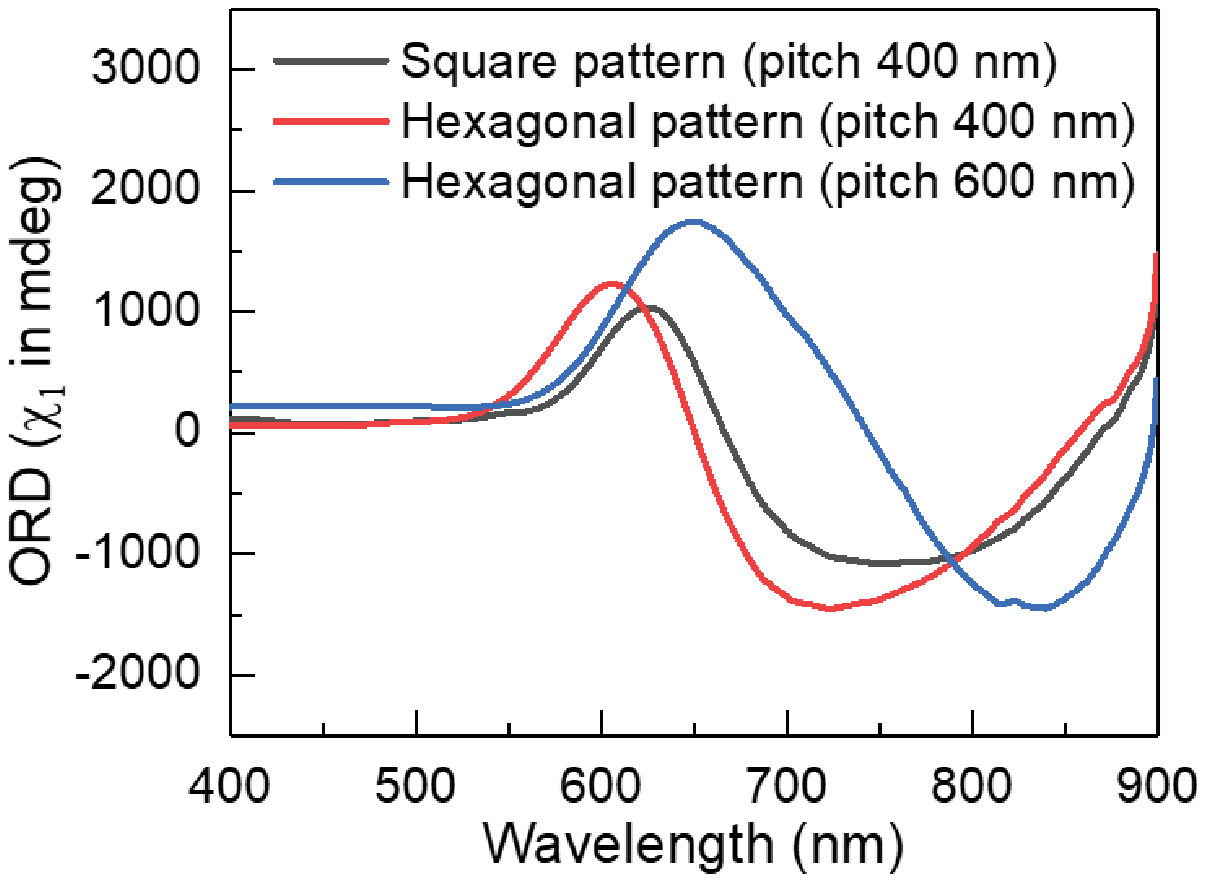}
		\label{ORD_p}
	\end{subfigure}
	
	\caption{The chiroptical properties of GNP plates with respect to the different patterns with the same size. The pitch means the gap between adjacent GNPs.}\label{pattern}
\end{figure}

We first explain some important chiroptical properties of GNP that vary the polarization of visible light in this section. The chiroptical properties of GNPs are the CD and ORD, which are the differential absorption and refraction of LCP and RCP light \cite{Lee:2018}. The differential refraction caused by the GNPs results in differential phase retardation between the incident LCP and RCP light. A GNP plate, which can be fabricated with a low cost, is made by judiciously stacking GNPs with different types and sizes \cite{Lee:2018}. This elaborated arrangement of GNPs makes it possible to precisely control the amplitudes and phases of outgoing LCP and RCP light. In the CP domain, the chiroptical properties of GNP plate for the polarized incident light can be represented as
\begin{equation}\label{eq1}
\mathbf{N}=\begin{bmatrix}
\sqrt{1-a_L} & 0\\
0 & \sqrt{1-a_R}e^{j\Delta\varphi}
\end{bmatrix}=
\begin{bmatrix}
\sqrt{\bar{a}_L} & 0\\
0 & \sqrt{\bar{a}_R}e^{j\Delta\varphi}
\end{bmatrix},
\end{equation}
where $a_L$ and $a_R$ are the absorption factors (i.e., $\bar{a}_L$ and $\bar{a}_R$ are the transmittance) for the LCP and RCP light with the range of $[0,1]$, and $\Delta\varphi$ is the difference of phase retardation between the LCP and RCP light. Figs. \ref{size} and \ref{pattern}, which are obtained by measuring the visible light passing through the GNP plates with the same procedure in \cite{Lee:2018}, show the CD and ORD for GNP plates with distinct sizes of GNPs and patterns of GNP plate. It is clear from the figures that, by using different sizes of GNPs and patterns of GNP plate, the CD and ORD significantly change with the wavelength of visible light \cite{Kim:2022}.

Next, we transform the GNP properties specified in the Stokes vector to the Jones vector to handle the VLC signals in the CP domain. Note that the Jones vector provides a straightforward representation for signals, but just for the fully polarized light while the Stokes vector spans the space of all the possible polarization states of electromagnetic radiation including unpolarized and partially polarized light as well as the fully polarized light. Thus, we also use the Stokes vector representation to describe the unpolarized transmit signal radiated from LEDs in Section \ref{sec_II}.\ref{eff_ch}.

The Stokes vector, which is associated with the intensity and polarization ellipse parameters of light, indicates the polarization state of electromagnetic field. The Jones vector $\begin{bmatrix} E_x\,\, E_y \end{bmatrix}^{\mathrm{T}}$ represents the polarization of light as transverse waves. The Stokes vector has the relation with the Jones vector as \cite{Eftimov:2009,Schaefer:2007}
\begin{equation}\label{eq2}
\begin{bmatrix}
I\\
Ip\,\mathrm{cos}(2\chi_1)\,\mathrm{cos}(2\chi_2)\\
Ip\,\mathrm{sin}(2\chi_1)\,\mathrm{cos}(2\chi_2)\\
Ip\,\mathrm{sin}(2\chi_2)
\end{bmatrix}=
\begin{bmatrix}
|E_x|^2+|E_y|^2\\
|E_x|^2-|E_y|^2\\
2\mathrm{Re}(E_xE_y^*)\\
-2\mathrm{Im}(E_xE_y^*)
\end{bmatrix},
\end{equation}
where $I$ is the total intensity of light, $p$ is the degree of polarization for the state of polarization, e.g., completely polarized, partially polarized, and unpolarized, $\chi_1$ and $\chi_2$ denote the ORD and CD, and $E_x$ and $E_y$ are the complex amplitudes of Jones vector in the linear domain, i.e., the Cartesian coordinates. By using this relation, we can define the transformer $\mathcal{J}\left\{\cdot\right\}$ from the Stokes vector into the Jones vector as
\begin{equation}\label{eqT}
\mathcal{J}\left\{\begin{bmatrix}
|E_x|^2+|E_y|^2\\
|E_x|^2-|E_y|^2\\
2\mathrm{Re}(E_xE_y^*)\\
-2\mathrm{Im}(E_xE_y^*)
\end{bmatrix}\right\}=\begin{bmatrix}
E_x\\
E_y
\end{bmatrix}.
\end{equation}
The LCP and RCP complex amplitudes $E_L$ and $E_R$ can be obtained using the transformation matrix $\bT_{\mathrm{L}\to\mathrm{C}}$ for $E_x$ and $E_y$ in (\ref{eqT}) as
\begin{equation}\label{eq3}
\begin{bmatrix}
E_L\\
E_R
\end{bmatrix}=\bT_{\mathrm{L}\to\mathrm{C}}
\begin{bmatrix}
E_x\\
E_y
\end{bmatrix},\quad
\bT_{\mathrm{L}\to\mathrm{C}}=\frac{1}{\sqrt{2}}
\begin{bmatrix}
1 & -j\\
1 & j
\end{bmatrix}.
\end{equation}

Note that the parameters related to the GNP properties in (\ref{eq1}) can be extracted from measurements. For example, by measuring the linearly polarized light with $0^{\circ}$ degree passed through a specific GNP, the LCP and RCP complex amplitudes of this GNP $E_L'$ and $E_R'$ are first obtained. With these values, the GNP properties in (\ref{eq1}) can be extracted from the relation \cite{Lee:2018}
\begin{align}\label{eq5}
\begin{bmatrix}
E_L'\\
E_R'
\end{bmatrix} = \bN\begin{bmatrix}
1\\
1
\end{bmatrix}.
\end{align}

\subsection{Effective Channel Modeling}\label{eff_ch}
In this subsection, we elaborate the signal and effective channel models including the effects of GNP plates and linear polarizers. In the scenarios of interest, $N_t$ LEDs that work as the transmitters transmit direct current (DC)-biased signals, which are composed with the intended symbol for Bob and the artificial noise to disguise Eve. The DC bias is necessary to obtain positive real-valued transmit signals after applying real-valued precoding techniques \cite{Qian:2014}. The real-valued symbol modulated transmit signals with the DC bias from the LEDs are represented as
\begin{align}
\bx &= \zeta P_{\mathrm{TX}}(I_{\mathrm{DC}}\mathbf{1}_{N_t} + \bs ),\notag\\
\bs&=\bw s_{\mathrm{I}} + \frac{1}{N_t}\bW_{\mathrm{a}} \bs_{\mathrm{a}},\label{eq4}
\end{align}
where $\zeta$ is the current-to-light conversion efficiency, $P_{\mathrm{TX}}$ is the optical power at the LED, $I_{\mathrm{DC}}\in \mathbb{R}_+$ is the DC bias to satisfy $|\bs|\preceq I_{\mathrm{DC}}\mathbf{1}_{N_t}$ with the precoded symbol $\bs$, and $\bw$ and $\bW_{\mathrm{a}}$ are the precoders for the intended symbol $s_{\mathrm{I}}$ and the artificial noise $\bs_{\mathrm{a}}$, respectively, with the constraint such that $\mathbb{E}\left\{| s_m |^2\right\}\le 1$. We randomly obtain the artificial noise $\bs_{\mathrm{a}}$ among symbols different from the intended symbol $s_{\mathrm{I}}$ within the constellation set. In (\ref{eq4}), $x_m$, the $m$-th component of $\bx$, represents the intensity of unpolarized transmit signal radiated from the $m$-th LED, which corresponds to $I=x_m$ and $p=0$ in (\ref{eq2}). Thus, $x_m$ is represented using the Stokes vector as \cite{Bickel:1985}
\begin{equation}\label{eq7}
\bx_{\mathrm{S},m}=\begin{bmatrix}
x_m\\
0\\
0\\
0
\end{bmatrix}.
\end{equation}

To increase the secrecy rate by the phase retardation effect of GNP plates, we employ commercially available linear polarizers that can control the phase retardation of light in the CP domain. The linear polarizer with the angle $\theta$ in the linear domain is well represented as the Jones matrix
\setcounter{equation}{7}
\begin{equation}\label{eqJ}
\bL_{xy}(\theta)=\begin{bmatrix}
\mathrm{cos}^2(\theta) & \mathrm{cos}(\theta)\mathrm{sin}(\theta) \\
\mathrm{cos}(\theta)\mathrm{sin}(\theta) & \mathrm{sin}^2(\theta)
\end{bmatrix}.
\end{equation}
We need the linear polarizer representation in the CP domain to derive the overall channel model with the GNP plates. From the fact that the linear polarization in the CP domain after the transformation by $\bT_{\mathrm{L}\to\mathrm{C}}$ equals to the transformation by $\bT_{\mathrm{L}\to\mathrm{C}}$ after linear polarization in the linear domain, the linear polarizer in the CP domain is derived as
\begin{equation}\label{eq6}
\bL_{LR}(\theta)=\bT_{\mathrm{L}\to\mathrm{C}}\bL_{xy}(\theta)\bT_{\mathrm{L}\to\mathrm{C}}^{-1}=\frac{1}{2}\begin{bmatrix}
1 & e^{-j2\theta}\\
e^{j2\theta} & 1
\end{bmatrix},
\end{equation}
which retards the RCP light by $2\theta$ with respect to the LCP light. We also need to obtain the M{\"u}ller matrix of linear polarizer in the linear domain to express the Stokes vectors of transmit signals having the unpolarized state as in (\ref{eq7}) since the Jones calculus only deals with the polarized light. The Jones matrix can be transformed into the M{\"u}ller matrix as \cite{Simon:1982}
\begin{equation}\label{eqM}
\bM=\bT_{\mathrm{J}\to\mathrm{M}}(\bJ\otimes\bJ^{\mathrm{*}})\bT_{\mathrm{J}\to\mathrm{M}}^{-1}
\end{equation}
with
\begin{align}
\bT_{\mathrm{J}\to\mathrm{M}} = \begin{bmatrix}
1 & 0 & 0 & 1\\
1 & 0 & 0 & -1\\
0 & 1 & 1 & 0\\
0 & j & -j & 0
\end{bmatrix},
\end{align}
where $\bJ$ is an arbitrary Jones matrix. Using the Jones matrix of linear polarizer in (\ref{eqJ}) and the transformation in (\ref{eqM}), the M{\"u}ller matrix of linear polarizer with the angle $\theta$ in the linear domain is given as
\begin{align}\label{eq10}
\bL_{xy}^{\mathrm{M}}(\theta) = \frac{1}{2}\begin{bmatrix}
1 & \mathrm{cos}(2\theta) & \mathrm{sin}(2\theta) & 0\\
\mathrm{cos}(2\theta) & \mathrm{cos}^2(2\theta) & \mathrm{sin}(2\theta)\mathrm{cos}(2\theta) & 0\\
\mathrm{sin}(2\theta) & \mathrm{sin}(2\theta)\mathrm{cos}(2\theta) & \mathrm{sin}^2(2\theta) & 0\\
0 & 0 & 0 & 0
\end{bmatrix}.
\end{align}
Using (\ref{eq7}) and (\ref{eq10}), the transmit signal after passing through the linear polarizer with the angle $\theta_m$ at the $m$-th transmitter is denoted as
\begin{equation}
\bL_{xy}^{\mathrm{M}}(\theta_m)\bx_{\mathrm{S},m}=\frac{x_m}{2}\begin{bmatrix}
1\\
\cos(2\theta_m)\\
\sin(2\theta_m)\\
0
\end{bmatrix}.
\end{equation}

The wireless channel model consisting of GNP plates, path-loss by the VLC geometry, and linear polarizer at the receiver can be represented in the CP domain by taking the propagation delay into account as
\setcounter{equation}{13}
\begin{align}\label{eq8}
&\tilde{\bH}_C=\sum_{n=0}^{N_{\mathrm{NLOS}}}\frac{1}{2}\begin{bmatrix}
\bI_{N_t} & e^{-j2\theta_{C}}\bI_{N_t}\\
e^{j2\theta_{C}}\bI_{N_t} & \bI_{N_t}
\end{bmatrix}\notag\\
&\,\,\times\begin{bmatrix}
\bG_{C,n}\mathrm{diag}(e^{j\boldsymbol{\vartheta}_{C,n}})\bar{\bA}_{L,C,n} & \mathbf{0}_{N_t\times N_t} \\
\mathbf{0}_{N_t\times N_t} & \bG_{C,n}\mathrm{diag}(e^{j\boldsymbol{\vartheta}_{C,n}})\bar{\bA}_{R,C,n}
\end{bmatrix}
\end{align}
with
\begin{align}\label{eq9}
\bar{\bA}_{L,C,n}&=\mathrm{diag}(\bar{\ba}_{L,C,n}),\notag\\
\bar{\bA}_{R,C,n}&=\mathrm{diag}(\bar{\ba}_{R,C,n}\odot e^{j\boldsymbol{\Delta\varphi}_{C,n}}).
\end{align}
In (\ref{eq8}) and (\ref{eq9}), $C\in\left\{ B,E \right\}$ with $B$ and $E$ denoting Bob and Eve, $\theta_{C}$ is the linear polarizer angle at the receiver, the $0$-th path is for the LOS path, $N_{\mathrm{NLOS}}$ is the number of NLOS paths obtained through a single reflection, $\bG_{C,n}=\mathrm{diag}(\bg_{C,n})$ with $\bg_{C,n}=\left[\sqrt{g_{C,1n}},\cdots,\sqrt{g_{C,N_tn}}\right]^{\mathrm{T}}$ where $g_{C,mn}$ is the $n$-th path-loss $g_{mn}$ for $C$, $\boldsymbol{\vartheta}_{C,n}=[\vartheta_{C,1n},\cdots,\vartheta_{C,N_tn}]^{\mathrm{T}}$ with $\vartheta_{C,mn}$ denoting the phase of the $n$-th path due to the propagation delay for $C$, and $\bar{\ba}_{L,C,n}$, $\bar{\ba}_{R,C,n}$, and $\boldsymbol{\Delta\varphi}_{C,n}$ are the transmittance and the phase retardation differences for the LCP and RCP light of the $n$-th path toward $C$, which are obtained from (\ref{eq5}).

\setcounter{equation}{19}
\begin{figure*}[t]
	\centering
	\vspace{-\baselineskip}
	\begin{align}\label{eq13}
	y_C&=\frac{\eta}{16}\sum_{m=1}^{N_t}\sum_{n=0}^{N_{\mathrm{NLOS}}}g_{C,mn}x_m\biggl( \left| \sqrt{\bar{a}_{L,C,mn}}e^{j(\vartheta_{C,mn}-\theta_m)} + \sqrt{\bar{a}_{R,C,mn}}e^{j(\vartheta_{C,mn}-2\theta_C+\theta_m+\Delta\varphi_{C,mn})} \right|^2\notag\\
	&\qquad\qquad\qquad\qquad + \left| \sqrt{\bar{a}_{L,C,mn}}e^{j(\vartheta_{C,mn}+2\theta_C-\theta_m)} + \sqrt{\bar{a}_{R,C,mn}}e^{j(\vartheta_{C,mn}+\theta_m+\Delta\varphi_{C,mn})} \right|^2 \biggr) + z_C\notag\\
	&=\frac{\eta}{8}\sum_{m=1}^{N_t}\sum_{n=0}^{N_{\mathrm{NLOS}}}g_{C,mn}x_m\biggl(\bar{a}_{L,C,mn}+\bar{a}_{R,C,mn}+2\sqrt{\bar{a}_{L,C,mn}\bar{a}_{R,C,mn}}\mathrm{cos}(2\theta_{C}-2\theta_m-\Delta\varphi_{C,mn})\biggr)+z_C\notag\\
	&=\frac{\eta}{8}\sum_{m=1}^{N_t}h_{\mathrm{eff},C,m}x_m+z_C=\frac{\eta}{8}\bh_{\mathrm{eff},C}^{\mathrm{T}}\bx+z_C
	\end{align}
	\hrule
\end{figure*}

The received signals before arriving at the PD are
\setcounter{equation}{15}
\begin{equation}\label{eq11}
\begin{bmatrix}
\br_{L,C}\\
\br_{R,C}
\end{bmatrix}=\tilde{\bH}_C\begin{bmatrix}
[\bT_{\mathrm{L}\to\mathrm{C}}\mathcal{J}\left\{\bL_{xy}^{\mathrm{M}}(\theta_1)\bx_{\mathrm{S},1}\right\}]_L\\
\vdots\\
[\bT_{\mathrm{L}\to\mathrm{C}}\mathcal{J}\left\{\bL_{xy}^{\mathrm{M}}(\theta_{N_t})\bx_{\mathrm{S},N_t}\right\}]_L\\
[\bT_{\mathrm{L}\to\mathrm{C}}\mathcal{J}\left\{\bL_{xy}^{\mathrm{M}}(\theta_1)\bx_{\mathrm{S},1}\right\}]_R\\
\vdots\\
[\bT_{\mathrm{L}\to\mathrm{C}}\mathcal{J}\left\{\bL_{xy}^{\mathrm{M}}(\theta_{N_t})\bx_{\mathrm{S},N_t}\right\}]_R
\end{bmatrix},
\end{equation}
where $\mathcal{J}\left\{\cdot\right\}$ is the transformer defined in (\ref{eqT}), and $[\cdot]_{\left\{L,R\right\}}$ denotes the LCP or RCP component. In (\ref{eq11}), $\bT_{\mathrm{L}\to\mathrm{C}}\mathcal{J}\left\{ \bL_{xy}^{\mathrm{M}}(\theta_m)\bx_{\mathrm{S},m} \right\}$ is simply given as
\begin{equation}\label{eq19}
\bT_{\mathrm{L}\to\mathrm{C}}\mathcal{J}\left\{ \bL_{xy}^{\mathrm{M}}(\theta_m)\bx_{\mathrm{S},m} \right\} = \frac{\sqrt{x_m}}{2}\begin{bmatrix}
e^{-j\theta_m}\\
e^{j\theta_m}
\end{bmatrix}.
\end{equation}
Using (\ref{eq19}), the received signals from the $m$-th transmitter in the CP domain $r_{L,C,m}$ and $r_{R,C,m}$, which are the $m$-th components of $\br_{L,C}$ and $\br_{R,C}$ in (\ref{eq11}), before arriving at the PD can be represented as
\begin{align}\label{eq20}
r_{L,C,m}&=\sum_{n=0}^{N_{\mathrm{NLOS}}}\frac{\sqrt{g_{C,mn}x_m}}{4}\biggl( \sqrt{\bar{a}_{L,C,mn}}e^{j(\vartheta_{C,mn}-\theta_m)}\notag\\
&\quad\qquad\quad+\sqrt{\bar{a}_{R,C,mn}}e^{j(\vartheta_{C,mn}-2\theta_C+\theta_m+\Delta\varphi_{C,mn})} \biggr),\notag\\
r_{R,C,m}&=\sum_{n=0}^{N_{\mathrm{NLOS}}}\frac{\sqrt{g_{C,mn}x_m}}{4}\biggl( \sqrt{\bar{a}_{L,C,mn}}e^{j(\vartheta_{C,mn}+2\theta_C-\theta_m)}\notag\\
&\quad\qquad\quad+\sqrt{\bar{a}_{R,C,mn}}e^{j(\vartheta_{C,mn}+\theta_m+\Delta\varphi_{C,mn})}\biggr).
\end{align}
Since the PD can measure only the intensity of signals \cite{fuada:2017}, the received signal at the PD is represented as
\begin{equation}\label{eq12}
y_C=\eta\sum_{m=1}^{N_t}\left(|r_{L,C,m}|^2+|r_{R,C,m}|^2\right)+z_C,
\end{equation}
where $\eta$ is the PD's responsivity, and $z_C\sim\mathcal{N}(0,\sigma_{\mathrm{t}}^2)$ is the thermal noise at the PD \cite{Komine:2004}. Using (\ref{eq20}) for (\ref{eq12}), the received signal at the PD is shown in (\ref{eq13}) at the top of this page, where $h_{\mathrm{eff},C,m}$ is given as
\setcounter{equation}{20}
\begin{align}\label{eq14}
h_{\mathrm{eff},C,m}&=\sum_{n=0}^{N_{\mathrm{NLOS}}}g_{C,mn}\biggl(\bar{a}_{L,C,mn}+\bar{a}_{R,C,mn}\notag\\
&\,+2\sqrt{\bar{a}_{L,C,mn}\bar{a}_{R,C,mn}}\mathrm{cos}(2\theta_{C}-2\theta_m-\Delta\varphi_{C,mn})\biggr),
\end{align}
which is the effective channel including the effects of path-loss, linear polarizers, and GNP plates. The received signal model in (\ref{eq13}) clearly shows that the phase $\boldsymbol{\vartheta}_{C,mn}$ due to the propagation delay in (\ref{eq8}) has no impact on the received signal at the PD. The final received signal model after eliminating the DC bias is represented as
\begin{align}
&\tilde{y}_C=\rho\bh_{\mathrm{eff},C}^{\mathrm{T}}\left(\bw s+\frac{1}{N_t}\bW_{\mathrm{a}} \bs_{\mathrm{a}}\right)+z_C,
\end{align}
where $\rho=\frac{\eta\zeta P_{\mathrm{TX}}}{8}$.

\section{Precoder Design and Linear Polarizer Angle Optimization}\label{prec_angle}
\subsection{Precoder design}\label{prec_design}
While some works, e.g., \cite{Cho:2020,Mostafa:2015}, are based on the assumption that the transmitters know the information of location and channel of Eve, we design the precoders $\bw$ and $\bW_{\mathrm{a}}$ with respect to the effective channel of Bob $\bh_{\mathrm{eff},B}$ since the information of location and channel of Eve cannot be known at the transmitters. To maximize the SINR of Bob, the precoders $\bw$ and $\bW_{\mathrm{a}}$ for the intended symbol and the artificial noise are chosen as
\setcounter{equation}{22}
\begin{align}\label{eq24}
\bw&=\frac{ \bh_{\mathrm{eff},B} }{ \|\bh_{\mathrm{eff},B} \| },\notag\\
\bW_{\mathrm{a}}&=\bI_{N_t} - \bh_{\mathrm{eff},B}\bh_{\mathrm{eff},B}^{\dagger},
\end{align}
where $\bW_{\mathrm{a}}$ is the orthogonal projection matrix with respect to $\bh_{\mathrm{eff},B}$. From (\ref{eq24}), it is clear that the maximum ratio transmission (MRT) precoder is used for the intended symbol to maximize the received power by aligning the phases of $\bh_{\mathrm{eff},B}$ as
\begin{equation}
\bh_{\mathrm{eff},B}^{\mathrm{T}}\bw = \| \bh_{\mathrm{eff},B} \|,
\end{equation}
which is also called as matched filter or coherent combining. The ZF precoder is used for the artificial noise, which projects the artificial noise on the nullspace of $\bh_{\mathrm{eff},B}$ to minimize its impact to Bob while the artificial noise works as interference for Eve as
\begin{align}
\bh_{\mathrm{eff},B}^{\mathrm{T}}\bW_{\mathrm{a}} &= \mathbf{0}^{\mathrm{T}},\notag\\ \bh_{\mathrm{eff},E}^{\mathrm{T}}\bW_{\mathrm{a}} &= \bh_{\mathrm{eff},E}^{\mathrm{T}} - \bh_{\mathrm{eff},E}^{\mathrm{T}}\bh_{\mathrm{eff},B}\bh_{\mathrm{eff},B}^{\dagger}\neq \mathbf{0}^{\mathrm{T}}.
\end{align}

\subsection{Linear polarizer angle optimization}
The SINRs of Bob and Eve are highly affected from the angles of linear polarizers. In the angle optimization, we consider the effective channels to verify the feasibility of GNP plates in the wiretapping scenarios. We first optimize the linear polarizer angles at the LEDs $\theta_m$ to minimize the effective channel of Eve. It will be clear, however, that the optimized linear polarizer angles are independent of the instantaneous effective channel of Eve, which makes the proposed solution highly practical. Then, the linear polarizer angle at Bob $\theta_B$ is obtained to maximize the effective channel of Bob with the optimized angles at the transmitters.

Under the use of precoders $\bw$ and $\bW_{\mathrm{a}}$, we need to obtain optimized linear polarizer angles at the transmitters $\boldsymbol{\theta}$ and Bob $\theta_B$ to maximize the secrecy rate. The SINRs at Bob and Eve can be represented as
\begin{align}\label{eq37}
&\mathrm{SINR}_B=
\frac{\rho^2\bh_{\mathrm{eff},B}^{\mathrm{T}}\bw\bw^{\mathrm{T}}\bh_{\mathrm{eff},B}}{\sigma_{\mathrm{t}}^2}=\frac{ \rho^2\| \bh_{\mathrm{eff},B} \|^2 }{\sigma_{\mathrm{t}}^2},\notag\\
&\mathrm{SINR}_E= \frac{\rho^2\bh_{\mathrm{eff},E}^{\mathrm{T}}\bw\bw^{\mathrm{T}}\bh_{\mathrm{eff},E}}{\frac{\rho^2}{N_t^2}\bh_{\mathrm{eff},E}^{\mathrm{T}}\bW_{\mathrm{a}}\bW_{\mathrm{a}}^{\mathrm{T}}\bh_{\mathrm{eff},E}+\sigma_{\mathrm{t}}^2}\notag\\
&\quad=\frac{ N_t^2 }{ \begin{pmatrix} \frac{ \bh_{\mathrm{eff},E}^{\mathrm{T}}\left( \bI + \frac{\frac{N_t^2\sigma_{\mathrm{t}}^2}{\rho^2}}{\|\bh_{\mathrm{eff},E}\|^2}\bI \right)\bh_{\mathrm{eff},E} }{ \bh_{\mathrm{eff},E}^{\mathrm{T}}\bh_{\mathrm{eff},B}\bh_{\mathrm{eff},B}^{\mathrm{T}}\bh_{\mathrm{eff},E} } -1 \end{pmatrix} }=\frac{N_t^2}{\frac{1}{f(\bh_{\mathrm{eff},B},\bh_{\mathrm{eff},E})}-1},
\end{align}
where
\begin{equation}
f(\bh_{\mathrm{eff},B},\bh_{\mathrm{eff},E}) = \frac{ \bh_{\mathrm{eff},E}^{\mathrm{T}}\bh_{\mathrm{eff},B}\bh_{\mathrm{eff},B}^{\mathrm{T}}\bh_{\mathrm{eff},E} }{ \bh_{\mathrm{eff},E}^{\mathrm{T}}\left( \bI + \frac{\frac{N_t^2\sigma_{\mathrm{t}}^2}{\rho^2}}{\|\bh_{\mathrm{eff},E}\|^2}\bI \right)\bh_{\mathrm{eff},E} }.
\end{equation}
According to (\ref{eq37}), $\| \bh_{\mathrm{eff},B} \|^2$ should be maximized to enhance $\mathrm{SINR}_B$ while $f(\bh_{\mathrm{eff},B},\bh_{\mathrm{eff},E})$ needs to be minimized to degrade $\mathrm{SINR}_E$. Simultaneously maximizing $\|\bh_{\mathrm{eff},B}\|^2$ and minimizing $f(\bh_{\mathrm{eff},B},\bh_{\mathrm{eff},E})$, however, is not possible in general. Therefore, we take a more practical approach in the following. Note that the components of $\bh_{\mathrm{eff},E}$ are independent of each other due to the receiving mechanism of PD that measures each intensity of received signal as in (\ref{eq12}). Assuming the angle of linear polarizer at Eve as $\theta_E=0$ (because this information cannot be known at the transmitters), the optimization problem to minimize the effective channel of Eve is represented as
\begin{gather}
\underset{\theta_m}{\min}\,\, \|\bh_{\mathrm{eff},E}\|^2\notag\\
\mathrm{s.t.}\,\,-\frac{\pi}{2}\le\theta_m\le\frac{\pi}{2}\quad\forall m,\label{eq15}
\end{gather}
which is represented using (\ref{eq14}) as
\begin{align}
&\underset{\theta_1,\cdots,\theta_{N_t}}{\min}\,\,f_E(\theta_1,\cdots,\theta_{N_t})\notag\\ &\qquad=\sum_{m=1}^{N_t}\sum_{n=0}^{N_{\mathrm{NLOS}}}g_{E,mn}\mathrm{cos}(2\theta_m+\Delta\varphi_{E,mn})\notag\\
&\qquad\times\biggl( \mathrm{cos}(2\theta_m+\Delta\varphi_{E,mn})+\sqrt{\frac{\bar{a}_{L,E,mn}}{\bar{a}_{R,E,mn}}}+\sqrt{\frac{\bar{a}_{R,E,mn}}{\bar{a}_{L,E,mn}}} \biggr)\notag\\
&\qquad\qquad\qquad\quad\mathrm{s.t.}\,\,-\frac{\pi}{2}\le\theta_m\le\frac{\pi}{2}\quad\forall m.\label{eq16}
\end{align}
We first obtain stationary points of the objective function in (\ref{eq16}) by solving $f'_E(\theta_1,\cdots,\theta_{N_t})=0$ as
\begin{align}\label{eq22}
f'_E(\theta_1,\cdots,\theta_{N_t})&=-\sum_{m=1}^{N_t}\sum_{n=0}^{N_{\mathrm{NLOS}}}2g_{E,mn}\sin(2\theta_m+\Delta\varphi_{E,mn})\notag\\
&\times\biggl( 2\cos(2\theta_m+\Delta\varphi_{E,mn})+u_{E,mn} \biggr)=0,
\end{align}
where $u_{E,mn}=\sqrt{\frac{\bar{a}_{L,E,mn}}{\bar{a}_{R,E,mn}}}+\sqrt{\frac{\bar{a}_{R,E,mn}}{\bar{a}_{L,E,mn}}}>2$ is self-evident for $\bar{a}_{L,E,mn}\ne\bar{a}_{R,E,mn}$, which always holds for GNPs with the chiroptical properties. Therefore, the stationary points are the angles that satisfy $\sum_{n=0}^{N_{\mathrm{NLOS}}}\tilde{g}_{E,mn}(\theta_m)\mathrm{sin}(2\theta_m+\Delta\varphi_{E,mn})=0$ with $\tilde{g}_{E,mn}(\theta_m)=g_{E,mn}\biggl(2\cos(2\theta_m+\Delta\varphi_{E,mn})+u_{E,mn} \biggr)$. The solution, however, requires to have the exact value of the difference of phase retardation of Eve $\Delta\varphi_{E,mn}$, which is a function of the location of Eve. To have a practical, but suboptimal, solution, we set
\begin{equation}\label{eq17}
\boldsymbol{\theta}^*\approx\left(\frac{k\pi}{2}-\frac{\mathbb{E}\left\{\Delta\varphi_{E,mn}\right\}}{2}\right)\mathbf{1}_{N_t}=\theta^*\mathbf{1}_{N_t},
\end{equation}
since the $m$-th element of optimal solution $\boldsymbol{\theta}_{\mathrm{opt}}^*$ can be obtained from the $i$-th iteration satisfying $\| \boldsymbol{\theta}_{\mathrm{opt},m}^i - \boldsymbol{\theta}_{\mathrm{opt},m}^{i-1} \|<\epsilon$ with an arbitrary small value $\epsilon$ as
\begin{align}\label{eq177}
\theta_{\mathrm{opt},m}^i&=\frac{k\pi}{2} + \frac{1}{2}\angle \sqrt{ \frac{ \sum_{n=0}^{N_{\mathrm{NLOS}}}\tilde{g}_{E,mn}( \theta_{\mathrm{opt},m}^{i-1} )e^{-j\Delta\varphi_{E,mn}} }{ \sum_{n=0}^{N_{\mathrm{NLOS}}}\tilde{g}_{E,mn}( \theta_{\mathrm{opt},m}^{i-1} )e^{j\Delta\varphi_{E,mn}} } }\notag\\
&\stackrel{(a)}{\approx}\frac{k\pi}{2} + \frac{1}{2}\angle \sqrt{ \frac{ \tilde{g}_{E,m0}(\theta_{\mathrm{opt},m}^{i-1}) e^{-j\Delta\varphi_{E,m0}} }{ \tilde{g}_{E,m0}(\theta_{\mathrm{opt},m}^{i-1}) e^{j\Delta\varphi_{E,m0}} } }\notag\\
&=\frac{k\pi}{2} - \frac{\Delta\varphi_{E,m0}}{2}\notag\\
&\stackrel{(b)}{\approx} \frac{k\pi}{2} - \frac{\mathbb{E}\left\{\Delta\varphi_{E,mn}\right\}}{2},
\end{align}
where $k$ is an integer that satisfies the constraints in (\ref{eq16}) and $f''_E(\theta^*,\cdots,\theta^*)>0$. In (\ref{eq177}), $(a)$ is from the fact that the LOS component has much larger channel gain compared to other NLOS components \cite{Komine:2004,Lee:2011,Mapunda:2020}, and $(b)$ is due to the unknown location of Eve, i.e., assuming the difference of phase retardation $\Delta\varphi_{E,mn}$ lies in the range of $[\Delta\varphi_{\mathrm{L}},\Delta\varphi_{\mathrm{U}}]$ with the lower bound $\Delta\varphi_{\mathrm{L}}$ and the upper bound $\Delta\varphi_{\mathrm{U}}$, we set $\mathbb{E}\left\{\Delta\varphi_{E,mn}\right\}=\frac{\Delta\varphi_{\mathrm{L}}+\Delta\varphi_{\mathrm{U}}}{2}$ to minimize the difference in average sense between the suboptimal solution $\boldsymbol{\theta}^*$ and the optimal solution $\boldsymbol{\theta}_{\mathrm{opt}}^*$. In Section \ref{simul}, we verify that the gap between secrecy rates when using the suboptimal solution and the optimal one is negligible.

Because of the narrow range of $[\Delta\varphi_{\mathrm{L}},\Delta\varphi_{\mathrm{U}}]$ as shown in \cite{Lee:2018}, it is reasonable to assume
\begin{equation}
|k\pi|\gg | \mathbb{E}\left\{ \Delta\varphi_{E,mn} \right\} - \Delta\varphi_{E,mn} |\quad\textrm{for}\,\,\, k\ne 0 \quad \forall m,n.
\end{equation}
Then, the second derivative of $f_E(\theta_1,\cdots,\theta_{N_t})$ can be approximated as
\begin{align}
&f''_E(\theta^*,\cdots,\theta^*)=-\sum_{m=1}^{N_t}\sum_{n=0}^{N_{\mathrm{NLOS}}}g_{E,mn}\notag\\
&\times\biggl( 8\cos(2k\pi-2\mathbb{E}\left\{\Delta\varphi_{E,mn}\right\}+2\Delta\varphi_{E,mn})\notag\\
&\qquad\qquad\quad+4\cos(k\pi-\mathbb{E}\left\{\Delta\varphi_{E,mn}\right\}+\Delta\varphi_{E,mn})u_{E,mn} \biggr)\notag\\
&\approx -\sum_{m=1}^{N_t}\sum_{n=0}^{N_{\mathrm{NLOS}}}\biggl( 8+4(-1)^k u_{E,mn} \biggr).
\end{align}
Note that $k=1$ makes $f''_E(\theta^*,\cdots,\theta^*)>0$ due to $u_{E,mn}>2$ for $\bar{a}_{L,E,mn}\ne\bar{a}_{R,E,mn}$. Thus, a specific suboptimal linear polarizer angle can be obtained by plugging in $k=1$ into (\ref{eq17}) as
\begin{equation}\label{eq30}
\theta^*=\frac{\pi}{2}-\frac{ \mathbb{E}\left\{ \Delta\varphi_{E,mn} \right\} }{2}.
\end{equation}

\setcounter{equation}{39}
\begin{figure*}[t]
	\centering
	\vspace{-\baselineskip}
	\begin{align}\label{eq32}
	(a):\,\, -\sum_{m=1}^{N_t}\sum_{n=0}^{N_{\mathrm{NLOS}}}2g_{B,mn}\sin(\tilde{\theta}_B-\Delta\varphi_{B,mn})\left( 2\cos(\tilde{\theta}_B-\Delta\varphi_{B,mn})+u_{B,mn} \right)&=0\notag\\
	(b):\,\, A_1\sin(\tilde{\theta}_B)\cos(\tilde{\theta}_B)+A_2(\sin^2(\tilde{\theta}_B)-\cos^2(\tilde{\theta}_B))+A_3\sin(\tilde{\theta}_B)+A_4\cos(\tilde{\theta}_B)&=0\notag\\
	(c):\,\, -jA_1( (t-t^{-1})(t+t^{-1}) )-A_2( (t-t^{-1})^2 + (t+t^{-1})^2 ) -j2A_3(t-t^{-1}) + 2A_4(t+t^{-1})&=0\qquad\qquad\notag\\
	(d):\,\, (-jA_1-2A_2)t^4+(-j2A_3+2A_4)t^3+(j2A_3+2A_4)t+jA_1-2A_2&=0
	\end{align}
	\hrule
\end{figure*}

\begin{figure}
	\centering
	\includegraphics[width=1.0\columnwidth]{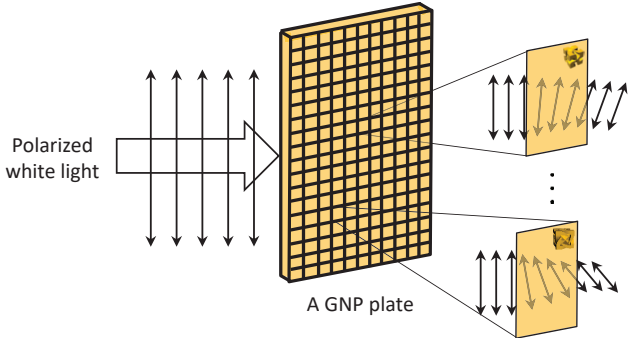}
	\caption{The channel variation effect by elaborated arrangement of GNPs.}\label{GNP_var}
\end{figure}

With $\theta^*$ obtained in (\ref{eq30}), the linear polarizer angle at Bob $\theta_B$ can be optimized as
\setcounter{equation}{35}
\begin{gather}
\underset{\theta_{B}}{\max}\,\, \|\bh_{\mathrm{eff},B}\|^2\notag\\
\mathrm{s.t.}\,\,-\frac{\pi}{2}\le\theta_{B}\le\frac{\pi}{2},\label{eq18}
\end{gather}
which can be represented in the same way of (\ref{eq16}) as
\begin{align}
\underset{\theta_B}{\max}\,\,f_B&(\theta_B)= \sum_{m=1}^{N_t}\sum_{n=0}^{N_{\mathrm{NLOS}}}g_{B,mn}\mathrm{cos}(\omega_{B,mn})\notag\\
&\times\biggl( \mathrm{cos}(\omega_{B,mn})+\sqrt{\frac{\bar{a}_{L,B,mn}}{\bar{a}_{R,B,mn}}}+\sqrt{\frac{\bar{a}_{R,B,mn}}{\bar{a}_{L,B,mn}}} \biggr) \notag\\
&\qquad\quad\quad\mathrm{s.t.}\,\,-\frac{\pi}{2}\le\theta_{B}\le\frac{\pi}{2}\label{eq26}
\end{align}
with
\begin{align}
\omega_{B,mn} &= 2\theta_B - 2\theta^* - \Delta\varphi_{B,mn}\notag\\
&=2\theta_B - \pi + \mathbb{E}\left\{ \Delta\varphi_{E,mn} \right\} - \Delta\varphi_{B,mn}.
\end{align}
Similar to (\ref{eq22}), the stationary points of $f_B(\theta_B)$ are first obtained by solving $f'_B(\theta_B)=0$ as
\begin{align}\label{eq29}
f'_B&(\theta_B)=-\sum_{m=1}^{N_t}\sum_{n=0}^{N_{\mathrm{NLOS}}}2g_{B,mn}\sin(\omega_{B,mn})\notag\\
&\times\biggl( 2\cos(\omega_{B,mn})+\sqrt{\frac{\bar{a}_{L,B,mn}}{\bar{a}_{R,B,mn}}}+\sqrt{\frac{\bar{a}_{R,B,mn}}{\bar{a}_{L,B,mn}}} \biggr)=0.
\end{align}
The solution of (\ref{eq29}) can be obtained by solving the quadratic equation as shown in (\ref{eq32}) at the top of this page with the following variables:
\setcounter{equation}{40}
\begin{align}
&\tilde{\theta}_B=2\theta_B-\pi+\mathbb{E}\left\{ \Delta\varphi_{E,mn} \right\}, \quad t=e^{j\tilde{\theta}_B},\notag\\
&u_{B,mn}=\sqrt{\frac{\bar{a}_{L,B,mn}}{\bar{a}_{R,B,mn}}}+\sqrt{\frac{\bar{a}_{R,B,mn}}{\bar{a}_{L,B,mn}}},\notag\\
&A_1=\sum_{m=1}^{N_t}\sum_{n=0}^{N_{\mathrm{NLOS}}}g_{B,mn}(2\cos^2(\Delta\varphi_{B,mn})-2\sin^2(\Delta\varphi_{B,mn})),\notag\\
&A_2=\sum_{m=1}^{N_t}\sum_{n=0}^{N_{\mathrm{NLOS}}}2g_{B,mn}\sin(\Delta\varphi_{B,mn})\cos(\Delta\varphi_{B,mn}),\notag\\
&A_3=\sum_{m=1}^{N_t}\sum_{n=0}^{N_{\mathrm{NLOS}}}g_{B,mn}\cos(\Delta\varphi_{B,mn})u_{B,mn},\notag\\
&A_4=-\sum_{m=1}^{N_t}\sum_{n=0}^{N_{\mathrm{NLOS}}}g_{B,mn}\sin(\Delta\varphi_{B,mn})u_{B,mn}.
\end{align}
In (\ref{eq32}), the equation (b) is obtained from (a) using the difference formulas of trigonometric functions $\sin(\alpha-\beta)=\sin\alpha\cos\beta-\cos\alpha\sin\beta$ and $\cos(\alpha-\beta)=\cos\alpha\cos\beta+\sin\alpha\sin\beta$, and the equation (c) is derived from (b) by the definitions of trigonometric functions in terms of the complex exponentials as $\sin\gamma = \frac{e^{j\gamma}-e^{-j\gamma}}{j2}$ and $\cos\gamma = \frac{e^{j\gamma}+e^{-j\gamma}}{2}$. After straightforward, but tedious derivations using the Cardano and Ferrari formulas \cite{Fujii:2013}, the closed-form expression for the quadratic equation (\ref{eq32}), $t^*$, can be obtained, which is omitted because of its tedious expression. With $t^*$, we have $\tilde{\theta}_B^*=\angle t^*$. Then, the linear polarizer angle at Bob can be set as
\begin{equation}\label{eq28}
\theta_B^*=\frac{ \pi - \mathbb{E}\left\{ \Delta\varphi_{E,mn} \right\} + \tilde{\theta}_B^* }{2},
\end{equation}
which satisfies the constraint in (\ref{eq26}) and $f''_B(\theta_B^*)<0$.

\subsection{Analysis of proposed design}
As shown in the previous subsections, the proposed GNP-empowered VLC system does not require any knowledge of Eve's CSI or location compared to previous VLC systems \cite{Cho:2020,Mostafa:2015}. While \cite{Mostafa:2014} also adopted the artificial noise, as in our approach, in addition to the ZF precoder without any knowledge of Eve's CSI or location, \cite{Mostafa:2014} may not work well when Eve is located near Bob since the effective channels of Bob and Eve become quite similar. In our proposed GNP-empowered VLC system, however, because the GNP plates at the transmitters can make the channels to all directions different using their chiroptical properties, i.e., the channels vary significantly even with small location variation thanks to the elaborated arrangement of GNPs as in Fig. \ref{GNP_var}, exploiting the GNP plates would be a good solution to increase the security at all possible locations.

\begin{figure}
	\centering
	\includegraphics[width=1.0\columnwidth]{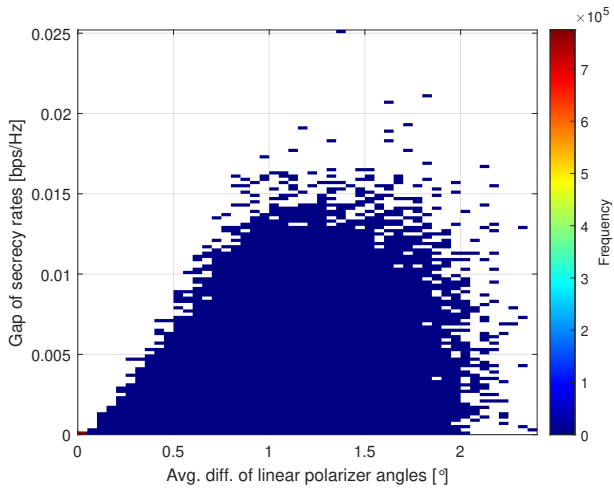}
	\caption{The gap between secrecy rates by the suboptimal solution in (\ref{eq17}) and the optimal one.}\label{diff_sec_rate}
\end{figure}

\begin{figure*}[h]
	\centering
	\begin{subfigure}{0.45\linewidth}
		\centering
		\includegraphics[width=\linewidth]{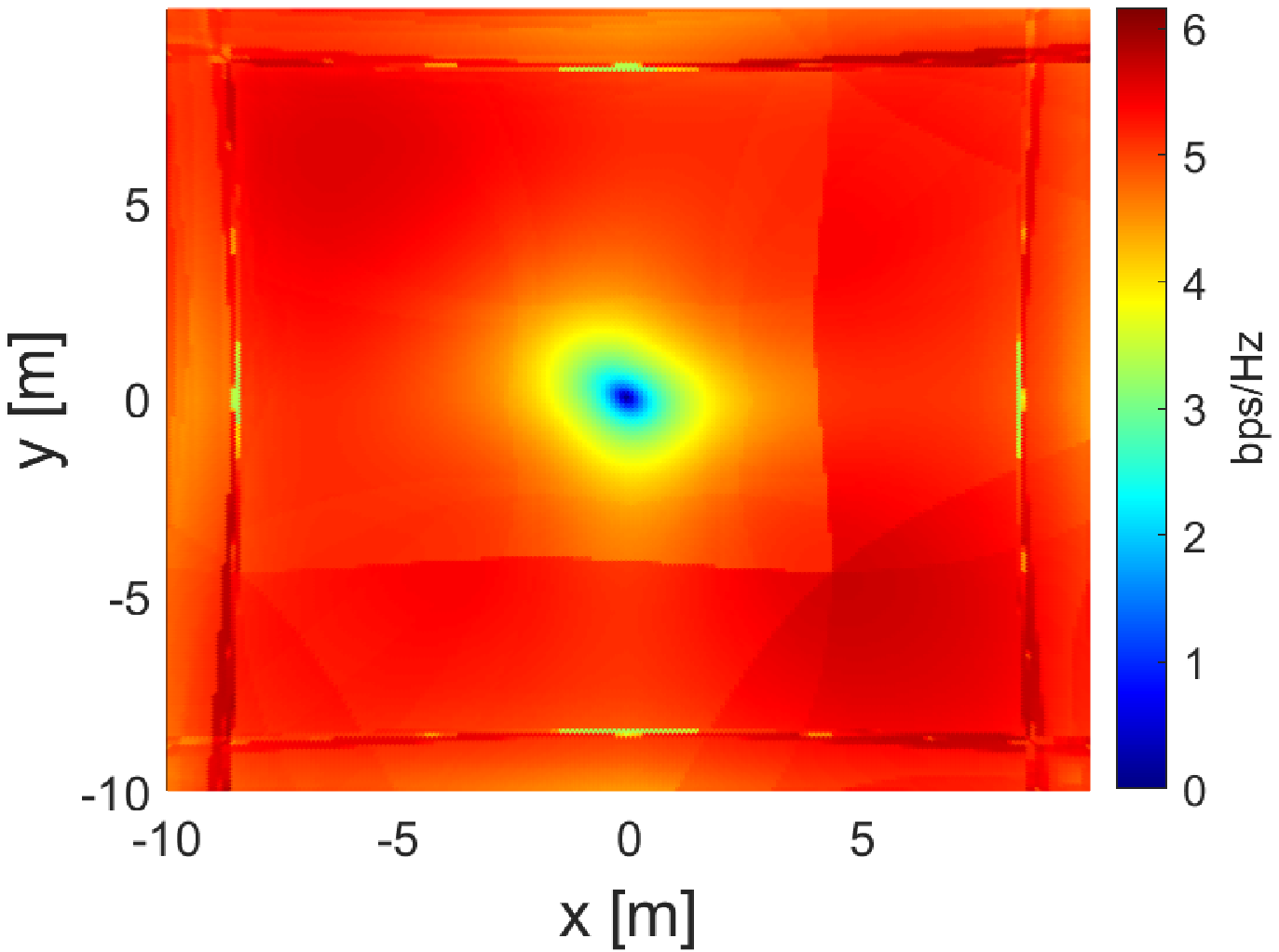}
		\caption{}
		\label{comp_scheme}
	\end{subfigure}
	\begin{subfigure}{0.45\linewidth}
		\centering
		\includegraphics[width=\linewidth]{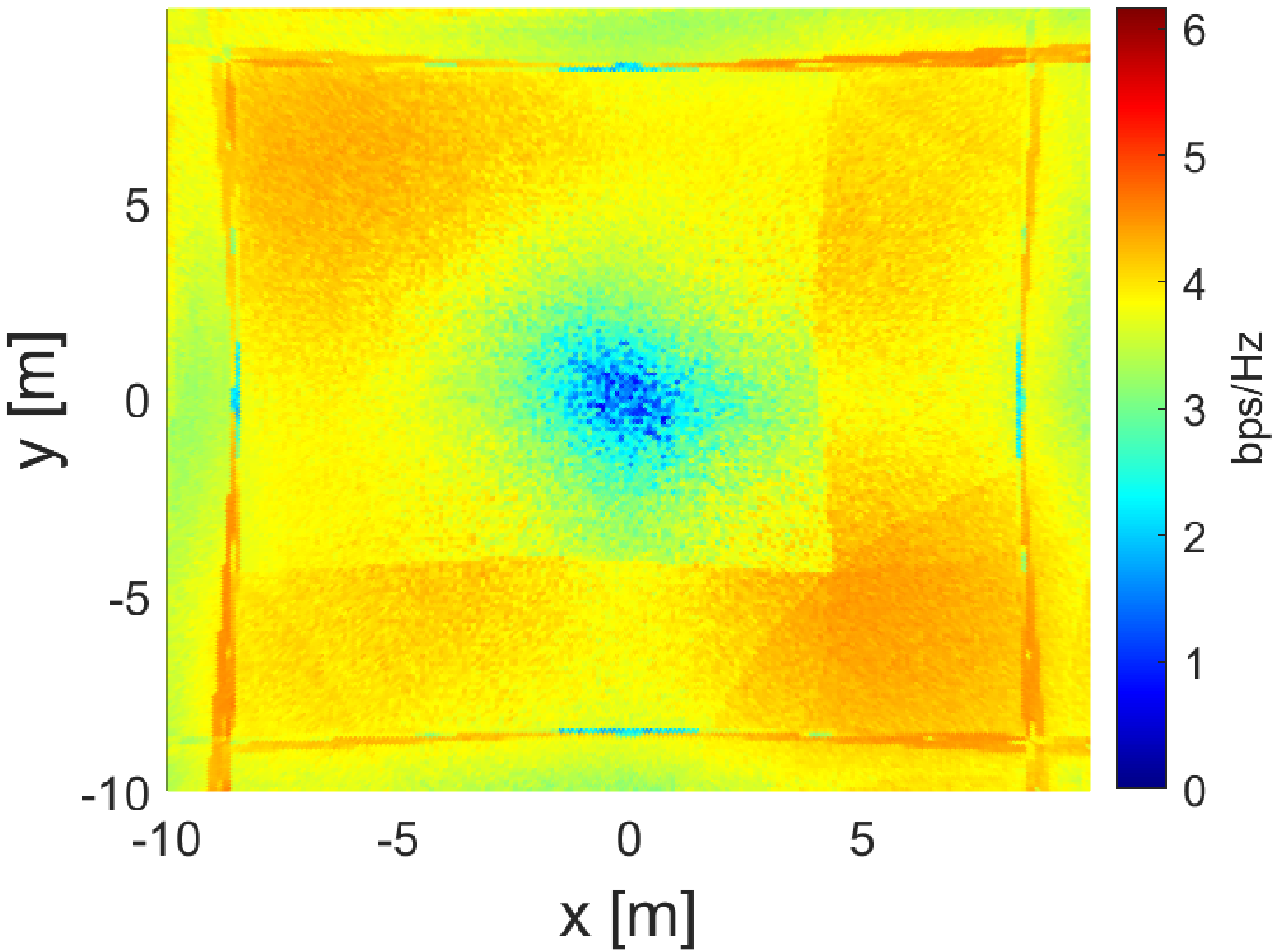}
		\caption{}
		\label{GNPb}
	\end{subfigure}
	\begin{subfigure}{0.45\linewidth}
		\centering
		\includegraphics[width=\linewidth]{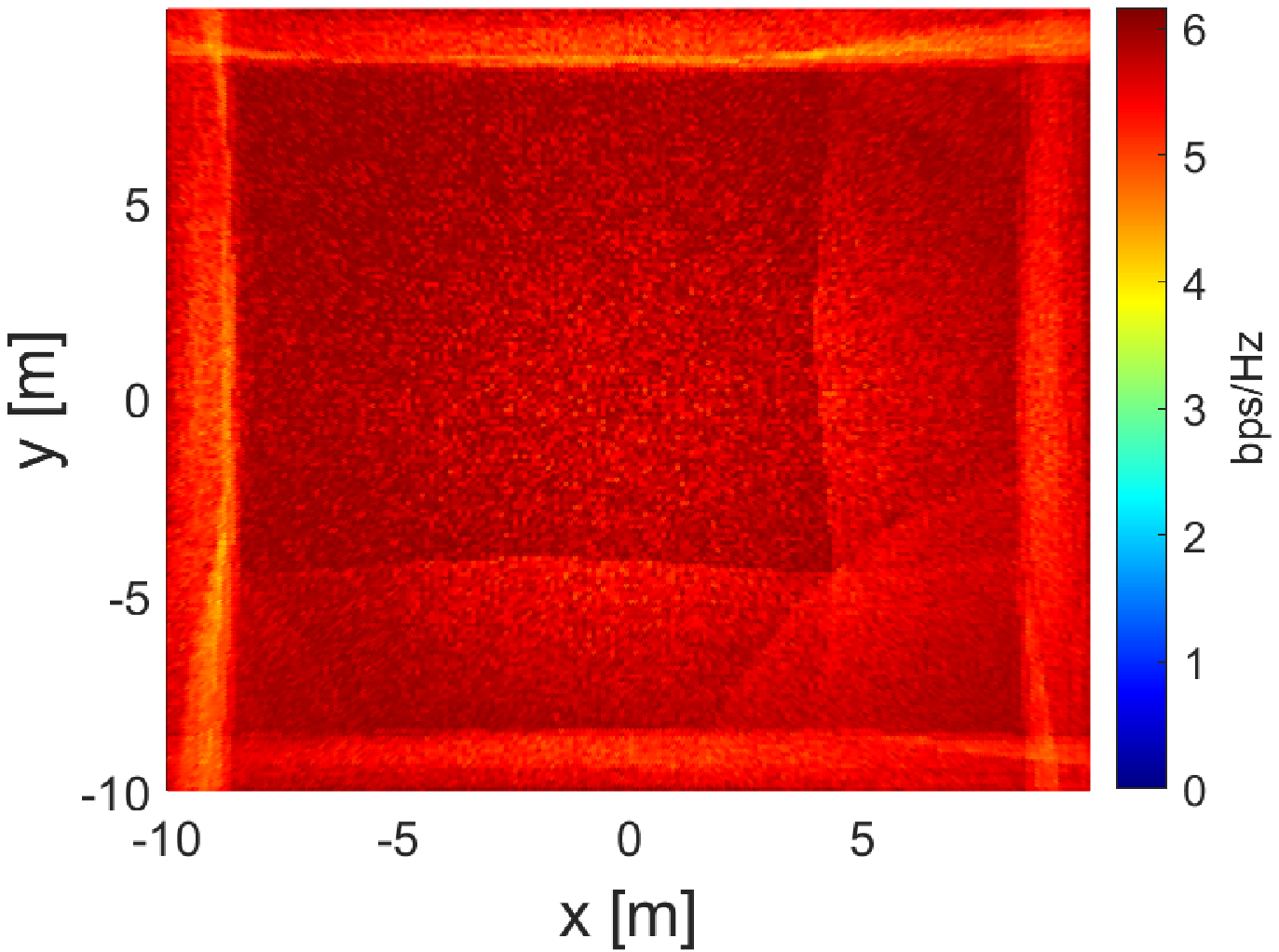}
		\caption{}
		\label{GNPa}
	\end{subfigure}
	\begin{subfigure}{0.45\linewidth}
		\centering
		\includegraphics[width=\linewidth]{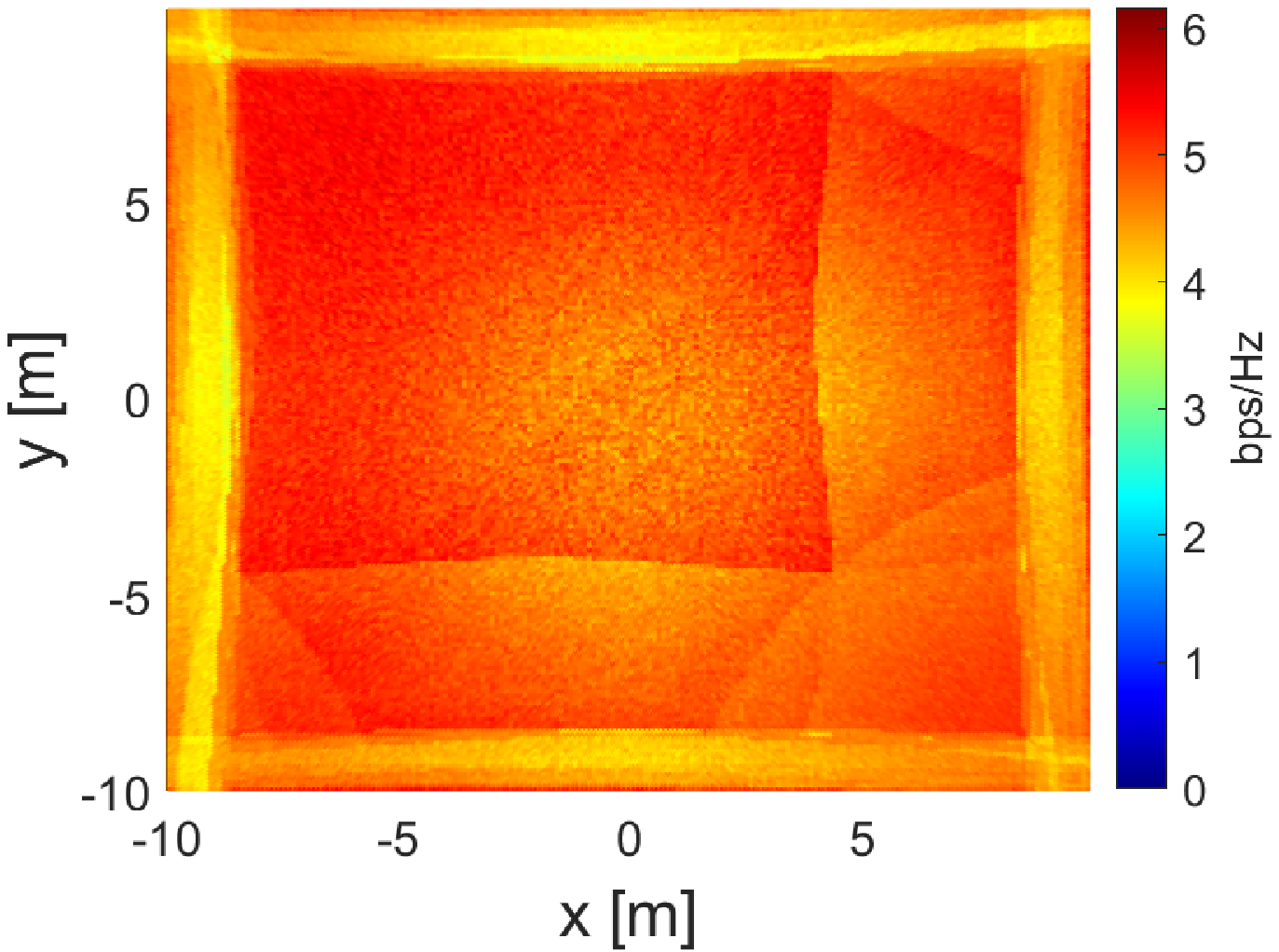}
		\caption{}
		\label{GNP_2pi_8}
	\end{subfigure}
	
	\caption{The secrecy rate performances of (a) the VLC system in \cite{Mostafa:2014}, (b) the GNP-empowered VLC system with $\theta_B$ in (\ref{eq28}) and $\theta_E=\theta_B$, (c) the GNP-empowered VLC system with $\theta_B$ in (\ref{eq28}) and $\theta_{E}=0$, and (d) the GNP-empowered VLC system with $\boldsymbol{\theta}=\frac{4\pi}{9}\mathbf{1}_{N_t}\neq \theta^*\mathbf{1}_{N_t}$, $\theta_B$ obtained by (\ref{eq28}) according to $\boldsymbol{\theta}$, and $\theta_E=0$.}\label{sec_rate}
\end{figure*}

\begin{figure}[h]
    \centering
    \includegraphics[width=\linewidth]{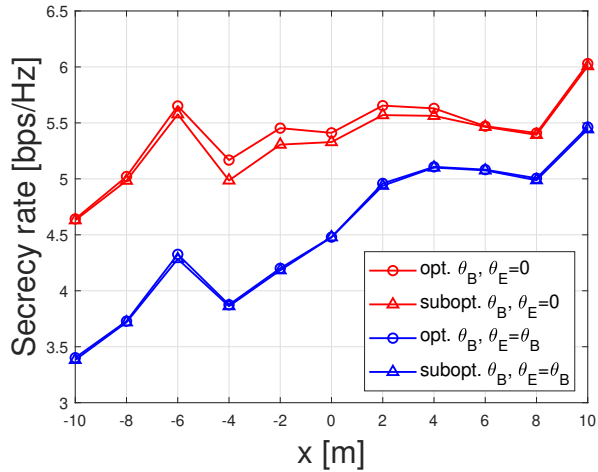}
    \caption{The secrecy rates depending on the location of Bob with ($x$ m, $0$ m, $1$ m) for the fixed location of Eve at ($-5$ m, $-5$ m, $1$ m).}\label{sec_position}
\end{figure}

\begin{figure*}[h]
	\centering
	\begin{subfigure}{0.45\linewidth}
		\centering
		\includegraphics[width=\columnwidth]{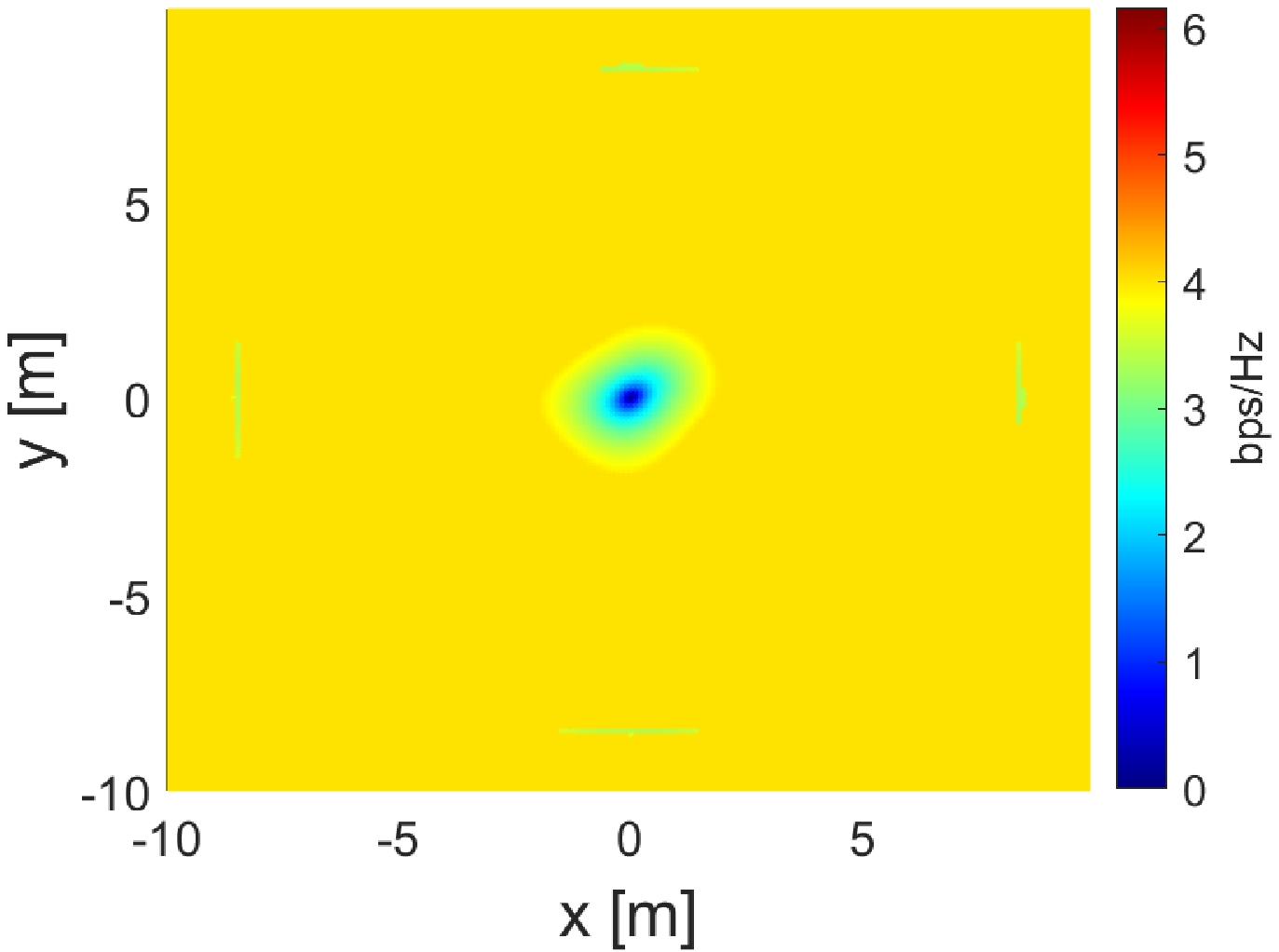}
		\caption{}
		\label{comp_scheme_multi}
	\end{subfigure}
	\begin{subfigure}{0.45\linewidth}
		\centering
		\includegraphics[width=\columnwidth]{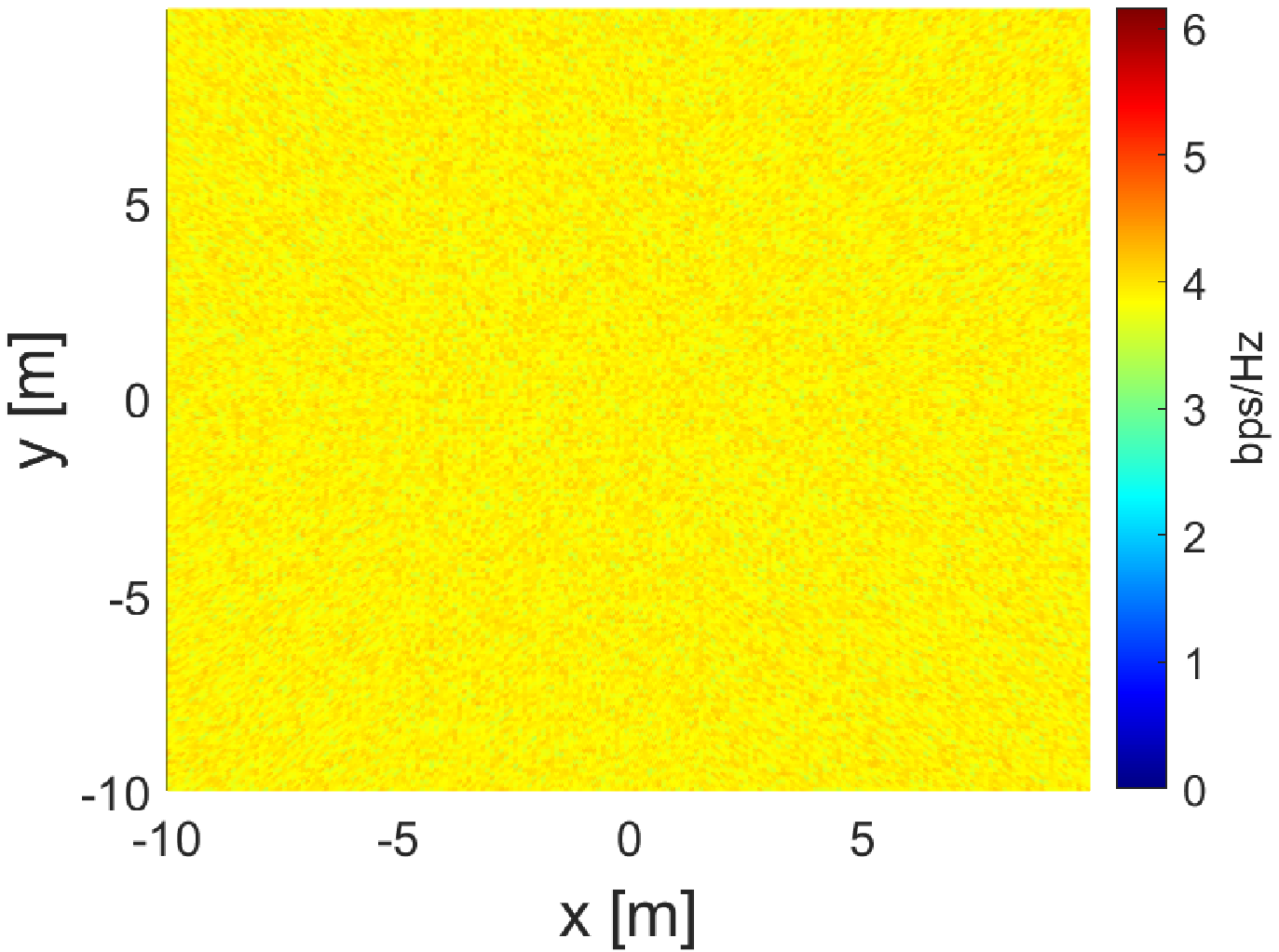}
		\caption{}
		\label{GNP_multi}
	\end{subfigure}
	
	\caption{The secrecy rate performances with three Eves. Two Eves are located at (-2, 0, 1) and (3, 4, 1) while the last Eve is randomly located. (a) the VLC system in \cite{Mostafa:2014}; (b) the GNP-empowered VLC system with $\theta_E=0$ for the randomly located Eve, $\theta_{E,\mathrm{fixed},1}=0$ for Eve located at (-2, 0, 1), and $\theta_{E,\mathrm{fixed},2}=\theta_B$ for Eve located at (3, 4, 1).}\label{multi_Eve}
\end{figure*}

\section{Simulation Results}\label{simul}
We present simulation results considering indoor eavesdropping scenarios to evaluate the proposed GNP-empowered VLC system in this section. Even though we demonstrate the performance of GNP-empowered VLC system by numerical simulations without an experimental setup, the realistic parameter setting from references assures the feasibility of GNP-empowered VLC system. The size of indoor room is $20$ m $\times$ $20$ m, the number and locations of LEDs are $N_t=4$ and ($\pm 2.2$, $\pm 2.2$, $4$) m, the number of NLOS paths is $N_{\mathrm{NLOS}}\in[204,295]$ depending on the location of Eve, and Bob and Eve are located in (0, 0, 1) m and ($\alpha$, $\beta$, 1) m for $\alpha$ and $\beta$ with the range of $-10\le \alpha,\beta\le 10$. The rest of simulation parameters are set as follows: the field of view (FOV) $\mathrm{FOV}=\frac{\pi}{2}$, the area of PD $A_{\mathrm{PD}}=1\,\mathrm{cm}^2$, the current-to-light conversion efficiency $\zeta=0.44$ W/A, the PD's responsivity $\eta=0.54$ A/W, the DC bias $I_{\mathrm{DC}}=3$ A, and the thermal noise variance $\sigma_{\mathrm{t}}^2=-133.8$ dBm \cite{Komine:2004,Cho:2021,Qian:2021}. From \cite{Lee:2018}, the lower and upper bounds of phase retardation difference are $\Delta\varphi_{\mathrm{L}}=0.6144$ and $\Delta\varphi_{\mathrm{U}}=0.8308$ in radian, and the absorption factors for the signals toward Bob and Eve randomly generated in the ranges of\footnote{Using multiple LEDs that have distinct phase retardation differences respect to the GNP plate at each transmitter, the assumption of less absorption factors for Bob than those for Eve seems to be reasonable since specific LEDs can be selected at each transmitter based on Bob's location such that the transmitted signals for Bob passed through the GNPs are absorbed the least.} $a_{L,B,mn}\in [0.1,0.11]$, $a_{R,B,mn}\in [0.25,0.26]$, $a_{L,E,mn}\in [0.1,0.4]$, and $a_{R,E,mn}\in [0.25,0.75]$. We set the optical power $P_{\mathrm{TX}}=10$ dBm.

\begin{figure*}[h]
	\centering
	\begin{subfigure}{0.45\linewidth}
		\centering
		\includegraphics[width=\linewidth]{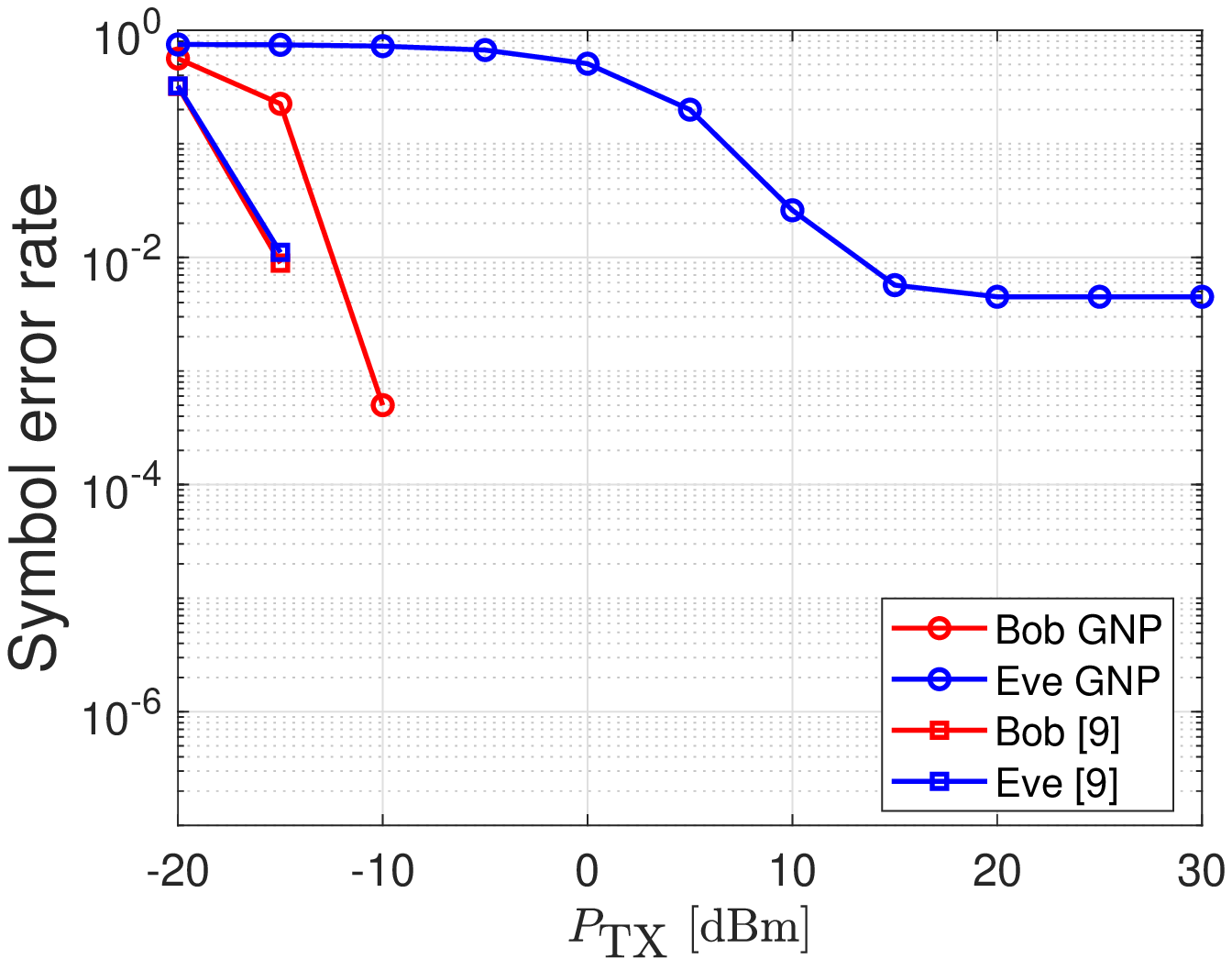}
		\caption{}
		\label{SER_a}
	\end{subfigure}
	\begin{subfigure}{0.45\linewidth}
		\centering
		\includegraphics[width=\linewidth]{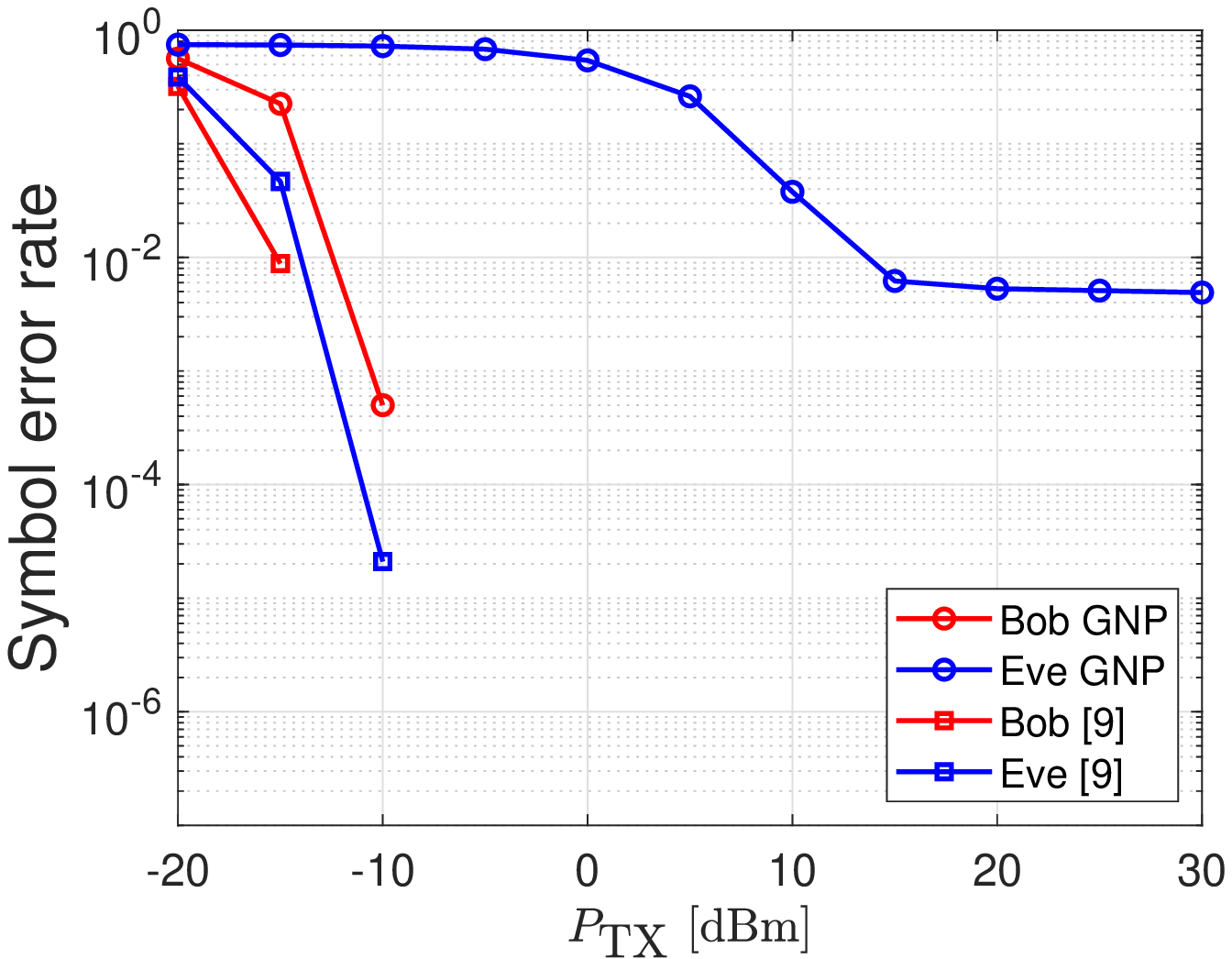}
		\caption{}
		\label{SER_b}
	\end{subfigure}
	\begin{subfigure}{0.45\linewidth}
		\centering
		\includegraphics[width=\linewidth]{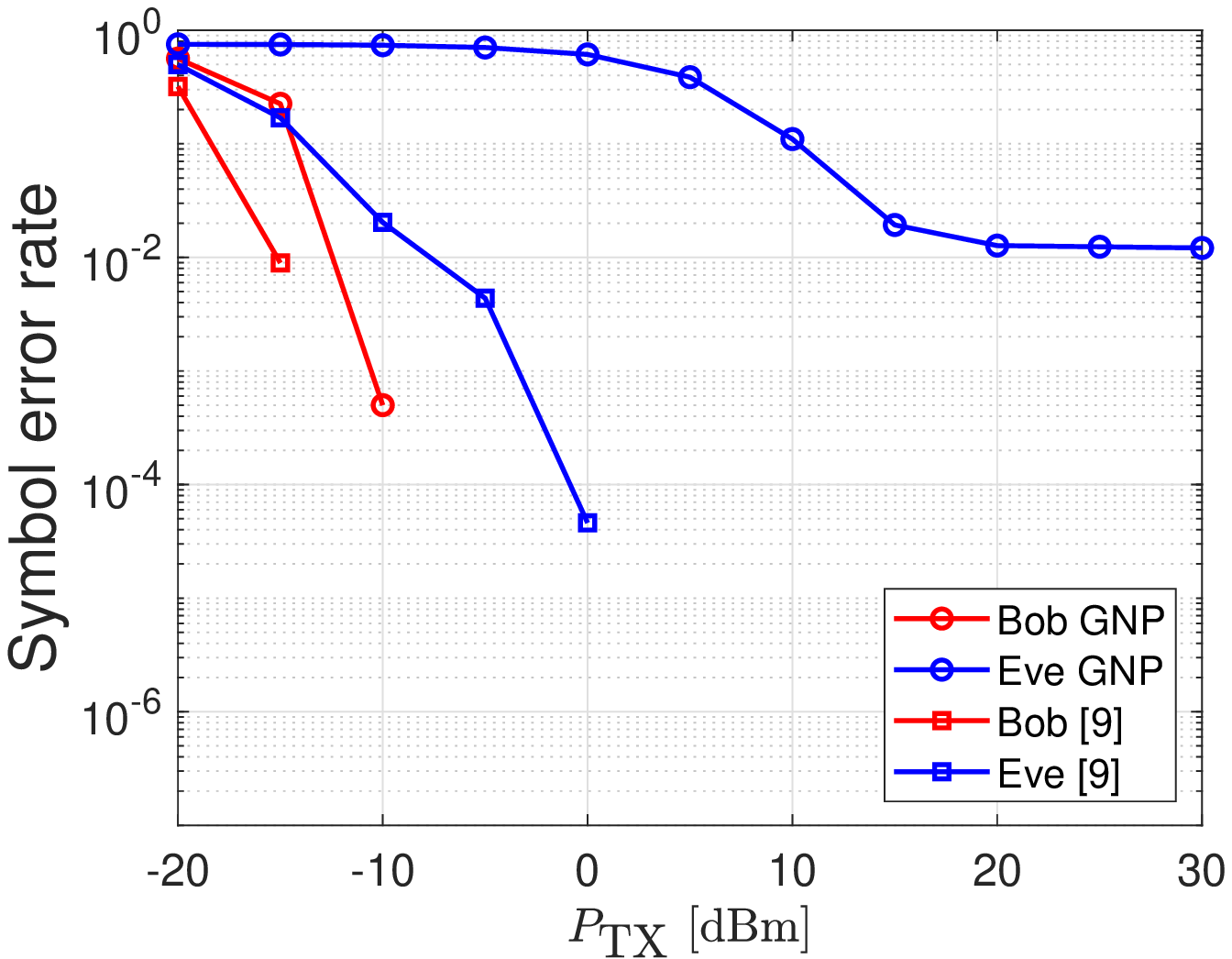}
		\caption{}
		\label{SER_c}
	\end{subfigure}
	\begin{subfigure}{0.45\linewidth}
		\centering
		\includegraphics[width=\linewidth]{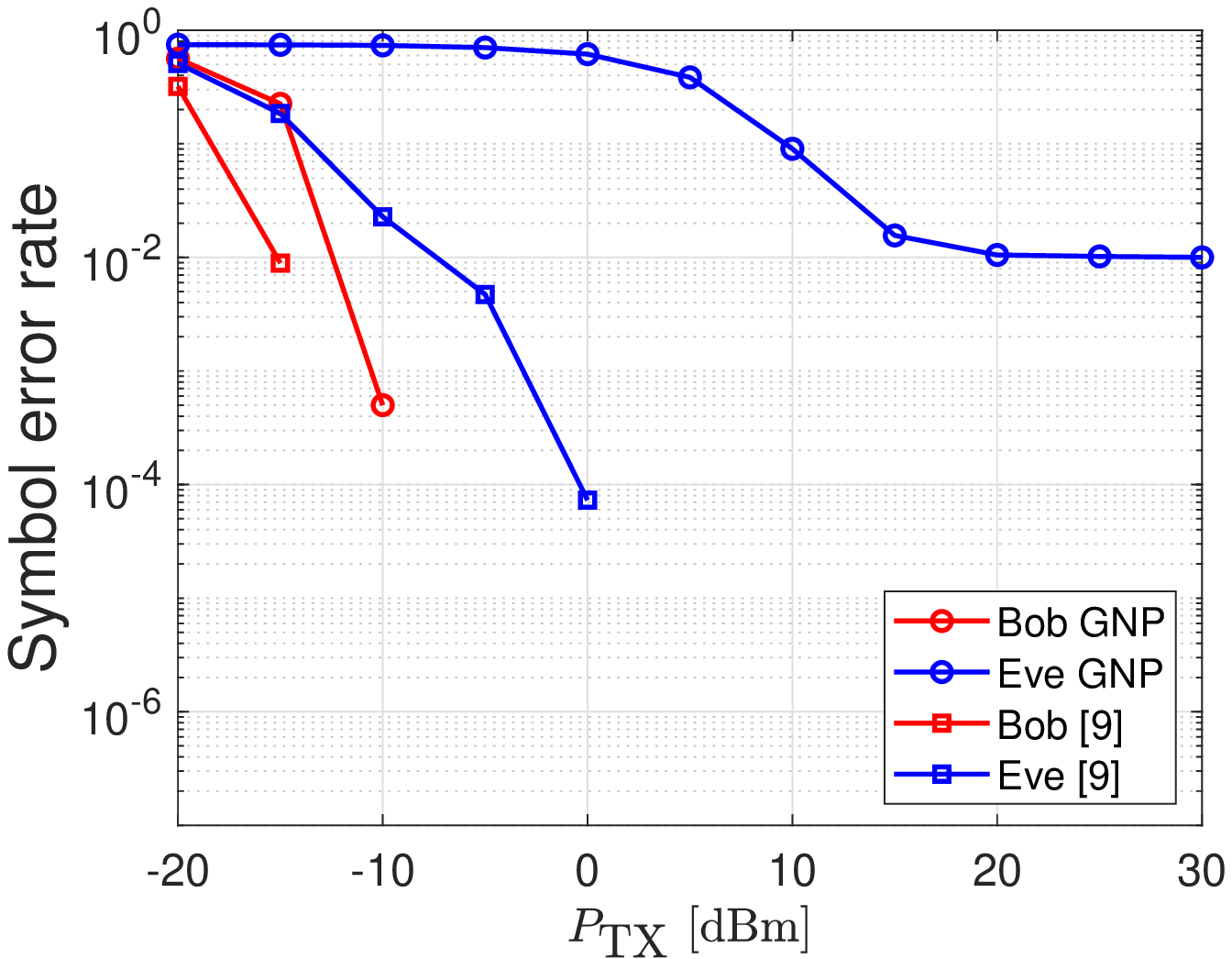}
		\caption{}
		\label{SER_d}
	\end{subfigure}
	
	\caption{The SER comparison for proposed technique with $\theta_E=0$ and \cite{Mostafa:2014} depending on Eve locations. (a) (1, 1, 1); (b) (2.5, 2.5, 1); (c) (5, 5, 1); (d) (7.5, 7.5, 1).}\label{SER}
\end{figure*}

We first consider the secrecy rate, i.e., the difference between the achievable rates of Bob and Eve defined as
\begin{align}
R_s=\mathrm{max}( R_B-R_E,0 )
\end{align}
with the achievable rates of Bob, $R_B$, and Eve, $R_E$, given as \cite{Elamassie:2021}
\begin{align}\label{eq23}
R_B&\approx\frac{1}{2}\mathrm{log}_2\left( 1 + \frac{e}{2\pi}\cdot\mathrm{SINR}_B \right),\notag\\
R_E&\approx\frac{1}{2}\mathrm{log}_2\left( 1 + \frac{e}{2\pi}\cdot\mathrm{SINR}_E \right),
\end{align}
where we simply assume the lower bound of capacity for VLC channel as the achievable rate since the closed-form expression for the exact capacity is unknown \cite{Yin:2017}.

In Fig. \ref{diff_sec_rate}, we first verify whether the suboptimal solution $\boldsymbol{\theta}^*$ in (\ref{eq17}) is indeed effective. Fig. \ref{diff_sec_rate} shows a histogram for the gap between secrecy rates depending on the averaged difference of linear polarizer angles $\mathbb{E}\left\{ |\boldsymbol{\theta}^*-\boldsymbol{\theta}_{\mathrm{opt}}^*| \right\}$ whose maximum value is about 2.4$^\circ$. The figure clearly shows that the gap is negligible even with the maximum polarizer angle difference, which shows the effectiveness of the suboptimal solution $\boldsymbol{\theta}^*$.

With the suboptimal linear polarizer angles at the transmitter $\boldsymbol{\theta}^*$, we now show the benefit of the GNP plates. Fig. \ref{sec_rate} shows the secrecy rates depending on the location of Eve for 1) the VLC system using the precoding techniques with the artificial noise in \cite{Mostafa:2014} and 2) the proposed GNP-empowered VLC system with different linear polarizer angles at Eve. The transmitters of VLC system in \cite{Mostafa:2014} use the MRT precoder for intended symbol and precoder constituted of bases in the null space of\footnote{While the proposed precoder design in Section \ref{prec_angle}.\ref{prec_design} is similar to \cite{Mostafa:2014}, our precoder exploits the effective channel $\bh_{\mathrm{eff},B}$ while \cite{Mostafa:2014} relies on the geometric channel $\tilde{\bg}_B$ to send the artificial noise.} $\tilde{\bg}_B$ as $\bW_{\mathrm{a}}\in\mathcal{N}(\tilde{\bg}_B)$ for artificial noise, where $\tilde{\bg}_B=\begin{bmatrix}\sum_{n=0}^{N_{\mathrm{NLOS}}}g_{B,1n},\cdots,\sum_{n=0}^{N_{\mathrm{NLOS}}}g_{B,N_t n}\end{bmatrix}$.

In Fig. \ref{comp_scheme}, it is possible for \cite{Mostafa:2014} to obtain high secrecy rates due to the artificial noise except the locations around Bob. This result is because the artificial noise works as strong interference to Eve. If Eve is close to Bob, however, the secrecy rate decreases significantly since the geometric channels of Bob and Eve become quite similar to each other. In Fig. \ref{GNPb}, the pessimistic case of proposed GNP-empowered VLC system when Eve can set the linear polarizer angle the same as Bob, which is usually not possible in practice, shows reduced secrecy rates compared to Fig. \ref{comp_scheme} due to the absorption factor and eavesdropping of $\theta_B$. On the contrary, in Fig. \ref{GNPa} when Eve does not know the polarizer angle of Bob and just set $\theta_E=0$, it is shown that the proposed GNP-empowered VLC system can provide high secrecy rates for all regions, even right next to Bob. Even though Eve tries to eavesdrop the intended symbol by estimating the integration of linear polarizer angles at transmitters and phase retardation for Eve with the exhaustive search, transmitters and Bob also can vary the linear polarizer angles to prevent eavesdropping. Fig. \ref{GNP_2pi_8} shows that we still have high secrecy rates even with a randomly selected $\boldsymbol{\theta}$ and $\theta_B$ obtained by (\ref{eq28}) according to $\boldsymbol{\theta}$. These results clearly show that it is possible to implement extremely secure VLC systems by exploiting the GNP plates.\footnote{We also verified the VLC system in \cite{Mostafa:2014} and proposed VLC system have similar secrecy rate performance tendency for a different number of LEDs.} The square lines at the edges of the room in Fig. \ref{sec_rate} are generated by the effects of NLOS paths, which can either enhance or attenuate the secrecy rates.

Fig. \ref{sec_position} shows the secrecy rates depending on the location of Bob with ($x$ m, $0$ m, $1$ m) for the fixed location of Eve at ($-5$ m, $-5$ m, $1$ m). We compare four cases for the linear polarizer angles at Bob and Eve ($\theta_B,\theta_E$). The gap between the secrecy rates by the suboptimal solution of $\theta_B$ and the optimal one is negligible as in Fig. \ref{diff_sec_rate} since it is determined by optimality of $\boldsymbol{\theta}$. The secrecy rates vary with the location of Bob and notably decrease for the unfavorable scenarios of $\theta_E=\theta_B$ while high secrecy rates are achieved thanks to the channel variation effect of GNP plates for $\theta_E\neq \theta_B$.

In Fig. \ref{multi_Eve}, we verify improved secrecy rates of proposed technique for a scenario with three Eves. Two Eves are located at (-2, 0, 1) and (3, 4, 1) while the last Eve is randomly located. The secrecy rate in Fig. \ref{multi_Eve} is obtained as
\begin{align}
R_s=\mathrm{max}\left( R_B-R_{E,\mathrm{max}},0 \right),
\end{align}
where $R_{E,\mathrm{max}}$ denotes the maximum of Eves' achievable rates. We assume Eves located at (-2, 0, 1) and (3, 4, 1) set the linear polarizer angles as $\theta_{E,\mathrm{fixed},1}=0$ and $\theta_{E,\mathrm{fixed},2}=\theta_B$, where one Eve can eavesdrop the linear polarizer angle of Bob, while the last Eve randomly located sets its angle as $\theta_E=0$. In Fig. \ref{comp_scheme_multi}, the VLC system in \cite{Mostafa:2014} has overall reduced secrecy rates compared to Fig. \ref{comp_scheme} due to Eves located at (-2, 0, 1) and (3, 4, 1) that are close to Bob. On the contrary, Fig. \ref{GNP_multi} shows high secrecy rates for all region compared to Fig. \ref{GNPa}, which means even Eves located near Bob are almost impossible to eavesdrop the intended symbol. These results show that the GNP-empowered VLC system outperforms the VLC system in \cite{Mostafa:2014} improving the secrecy rates even with multiple Eves. In particular, the high secrecy rates around Bob are due to the effect of channel variation by the GNP plates.

Fig. \ref{SER} shows the SER depending on Eve's locations where the symbol detection at Bob and Eve is performed by the maximum likelihood (ML) detection with the full CSI, where the intended symbol $s_{\mathrm{I}}$ and artificial noise $\bs_{\mathrm{a}}$ are randomly selected from the 4-pulse amplitude modulation (PAM) constellation. The SER is defined as
\begin{equation}
\textrm{SER}=\frac{1}{N_{\mathrm{I}}}\sum_{i=1}^{N_{\mathrm{I}}}\mathbb{I}\left( | s_{\mathrm{I},i} - \hat{s}_{\mathrm{I},i} | \right),
\end{equation}
where $\mathbb{I}(v)$ is the indicator function giving $1$ for $v\ne 0$ and $0$ otherwise, $N_{\mathrm{I}}$ is the number of iteration, and $(s_{\mathrm{I},i},\hat{s}_{\mathrm{I},i})$ are the true and detected intended symbols at the $i$-th iteration. In Fig. \ref{SER}, the SER plots clearly show the GNP-empowered VLC system indicated as ``Bob/Eve GNP'' shows much larger SER gap compared to the VLC system in \cite{Mostafa:2014} marked as ``Bob/Eve \cite{Mostafa:2014}''. The gap between the SERs of Bob is not large while the SERs for Eve show huge difference since it is much difficult for Eve to detect the intended symbol in the GNP-empowered VLC system. Because of the various arrangement of GNPs with different types, patterns, and sizes inside the GNP plates, the channels for all radiation directions are significantly different, making the artificial noise designed in Section \ref{sec_II}.\ref{eff_ch} highly effective to Eve. On the contrary, the eavesdropping in the high SNR region is still possible when using the technique of \cite{Mostafa:2014} even though Eve is located far from Bob as in Fig. \ref{SER_d}.

Note that due to the absorption of GNP plates, the SER performance of Bob in the GNP-empowered VLC system is degraded compared to that in the VLC system of \cite{Mostafa:2014}, and the enhanced physical layer security comes from the greater deterioration in the SER performance of Eve. While the proposed GNP-empowered VLC system makes it almost impossible for Eve to detect the transmitted symbols with the cost of Bob's SER performance, the degradation is negligible in practice since the practical power driving range of indoor LED bulbs is more than $30$ dBm \cite{Calderon:2015}. Moreover, the GNP plates may experience less absorption with more advanced fabricating techniques \cite{Lee:2022,Kim:2020}, making it possible to further improve the performance of proposed GNP-empowered VLC system.

The performance of proposed GNP-empowered VLC system is determined by mainly the linear polarizer angles at transmitters and Bob and the chiroptical properties of GNPs. To effectively enhance the physical layer security with the proposed system, however, there are several practical issues to be considered. The tilt angles of linear polarizers may occur the received power degradation by increasing the incident angle at the PD. The chiroptical properties of GNPs shared between transmitters and Bob can be varied by the location of Bob due to the elaborated arrangement of GNPs. Also, a receiver located in a specific location may be affected from multiple GNPs due to the multiple light sources deployed in different positions within each LED. These challenges should be properly addressed to make the proposed GNP-empowered VLC system practical.

\section{Conclusion}\label{conc}
GNPs are recently synthesized metamaterials that can differentially absorb and retard LCP and RCP light. In this paper, we proposed a secure GNP-empowered VLC system with linear polarizers. The transmitters are equipped with well-structured GNP plates, which consist of various types of GNPs, to prevent wiretapping by exploiting the chiroptical properties of GNP plate. A novel VLC channel model considering the GNP plates and linear polarizers was derived in the CP domain to obtain the received signal model. Then, the precoder design and linear polarizer angle optimization were performed to achieve the high secrecy rate. Numerical results showed that the use of GNP plates leads to a notable enhancement of the secrecy rates in the regions around Bob. Also, it was verified that the GNP plates make the symbol detection at Eve almost impossible even for high SNR region. The proposed GNP-empowered VLC system can be adopted to many wireless communication scenarios that require high level security, e.g., military and vehicular communications.

There are many possible future works for the GNP-empowered VLC system. Robust beamformers can be adopted for more practical scenario with imperfect channel information of Bob by limited feedback \cite{Mostafa:2016}. The angle diversity receiver having a wide field of view to receive signals from multiple directions also needs to be considered since the receiver is not always non-stationary. The GNP-empowered VLC system also can be adapted to WDM as well as spatial multiplexing to achieve higher data rates \cite{Chen:2022}.

\section*{Funding}This research was supported in part by the Challengeable Future Defense Technology Research and Development Program(912909601) of Agency for Defense Development in 2022, and in part by Institute of Information \& communications Technology Planning \& Evaluation (IITP) grant funded by the Korea government(MSIT) (No.2021-0-00847, Development of 3D Spatial Satellite Communications Technology).

\bibliography{refs_all}
  
\end{document}